\documentclass{article}

\usepackage{arxiv}

\usepackage[utf8]{inputenc} 
\usepackage[T1]{fontenc}    
\usepackage{hyperref}       
\usepackage{url}            
\usepackage{booktabs}       
\usepackage{amsfonts}       
\usepackage{microtype}      
\usepackage{tabularx} 
\usepackage{graphicx}
\usepackage[numbers]{natbib}
\usepackage{amsmath}

\usepackage{subcaption}

\title{Measuring the Predictability of Recommender Systems using Structural Complexity Metrics}

\author{
Andrés Abeliuk \\
Department of Computer Science\\
University of Chile\\
Center for Artificial Intelligence (CENIA)\\
Santiago, Chile \\
\And
Alfonso Valderrama \\
Department of Computer Science\\
University of Chile\\
Center for Artificial Intelligence (CENIA)\\
Santiago, Chile \\
\And
Simón Campos \\
Department of Computer Science\\
University of Chile\\
Santiago, Chile \\
\And
Marcelo Mendoza \\
Department of Computer Science\\
Pontificia Universidad Católica de Chile\\
Center for Artificial Intelligence (CENIA)\\
Santiago, Chile \\
}
\makeatletter
\renewcommand{\@date}{}
\renewcommand{\today}{}
\makeatother

\begin{document}
\maketitle

\begin{abstract}
Recommender Systems (RS) shape the filtering and curation of online content, yet we have limited understanding of how predictable their recommendation outputs are. We propose data-driven metrics that quantify the predictability of recommendation datasets by measuring the structural complexity of the user–item interaction matrix. High complexity indicates intricate interaction patterns that are harder to predict; low complexity indicates simpler, more predictable structures. 
We operationalize structural complexity via data perturbations, using singular value decomposition (SVD) to assess how stable the latent structure remains under perturbations. Our hypothesis is that random perturbations minimally affect highly organized data, but cause substantial structural disruption in intrinsically complex data. By analyzing prediction errors on perturbed interactions, we derive metrics that quantify this sensitivity at both the dataset and the interaction levels, yielding a principled measure of inherent predictability. 
Experiments on real-world datasets show that our structural complexity metrics correlate with the performance of state-of-the-art recommendation algorithms. We also demonstrate structure-aware data selection: in low-data settings, models trained on a carefully chosen subset of interactions with low structural perturbation error consistently outperform models trained on the full dataset. Thus, structural complexity serves both as a precise diagnostic of dataset complexity and as a principled foundation for efficient, data-centric training of RS.
\end{abstract}


\section{Introduction}

As the volume of available content continues to grow, recommender systems (RS) have become crucial to enhance user experiences in e-commerce, entertainment, social networks, and personalized content delivery. Among various recommendation paradigms, collaborative filtering (CF) stands out as a widely adopted approach that leverages users' preferences and behaviors to recommend items based on taste similarity~\cite{shi2014collaborative}. Understanding CF mechanisms remains challenging due to the large-scale, high-dimensional, and structurally intricate nature of user-item interaction datasets, which hinders the identification of informative patterns and characterization of underlying generative processes. Despite its practical success and the increasing use of interdisciplinary methodologies from data science, machine learning, and social sciences~\cite{chen2023bias}, CF still lacks a rigorous framework for quantifying the intrinsic predictability of recommendation datasets. Specifically, there is no principled measure of how learnable the underlying interaction patterns truly are, independent of any specific algorithm.

Predictability is a fundamental dimension of complexity in CF systems. Evaluating the predictability of user–item interaction data in CF–based RS can provide insight into the limitations of the algorithms and data used. A low predictability score indicates highly complex and uncertain user–item relationships, which can undermine the accuracy and robustness of the resulting recommendations. In contrast, a high predictability score indicates that the model effectively captures latent structures and user preference patterns, thereby producing more reliable and personalized recommendation outputs. Although predictability has been extensively studied in networks~\cite{lu2015toward}, time series~\cite{garland2014model}, and partially observed systems~\cite{abeliuk2020partially}, its systematic characterization in collaborative filtering remains, to the best of our knowledge, largely unexplored.

We introduce a data-driven structural complexity metric that quantifies the intrinsic predictability of recommendation datasets by measuring how well their interaction patterns withstand random perturbations. This complexity metric enables the assessment of an explainable structure and the characterization of the dataset's difficulty, independent of any specific algorithm.

Recent work has explored the relationship between dataset characteristics and recommendation performance. \citet{mcelfresh2022reczilla} proposed using meta-features, such as mean rating, Shannon entropy, and Gini coefficient, to select the most suitable algorithm for a given dataset. 
Our work takes a fundamentally different approach. Rather than predicting which algorithm will succeed, we have developed a metric that quantifies the inherent structural complexity of recommendation data, independent of any specific algorithmic approach. This enables dataset characterization and, as we demonstrate, structure-aware data selection for efficient model training.

We introduce two complementary model-agnostic complexity metrics for RS datasets, formalized and validated against the performance of state-of-the-art algorithms on real-world datasets. Our empirical evaluations demonstrate that both metrics strongly correlate with algorithmic accuracy, establishing their effectiveness for characterizing dataset predictability.
Furthermore, we show that this framework enables effective, structure-aware data selection: in low-data regimes, models trained on a carefully curated subset of interactions, selected for their low structural perturbation error, consistently outperform models trained on the full dataset. To favor reproducibility, the code and datasets used in our computations are freely available~\footnote{Experiment 1: \url{https://anonymous.4open.science/r/RS_Benchmark-6BB7}; Experiment 2: \url{https://anonymous.4open.science/r/RS-Structural-Complexity-1FA1}}.


This paper is structured as follows. Section \ref{sec:relwork} provides a review of the relevant literature. In Section \ref{sec:metrics}, we formally define and justify the two evaluation metrics examined in this study. Section \ref{sec:exps} describes the experimental methodology employed to assess these metrics on real-world datasets and reports the corresponding analytical results. Finally, Section \ref{sec:conc} presents conclusions, discusses the limitations of this study and outlines future work.

\section{Related work}
\label{sec:relwork}

To the best of our knowledge, this is the first study to tackle predictability and complexity in RS head-on. While prediction in RS can be framed as a special case of link prediction in graphs, our approach goes further, leveraging established metrics of the structural properties of complex networks to expose the limits of their predictability.

\subsection{Predictability in complex networks}

The study on link prediction by \citet{sun2020revealing} introduces an entropy-based metric to quantify predictability in complex networks. The authors define a measure of structural entropy by employing the lossless compression algorithm proposed in \cite{choi2012compression}, which encodes the network topology into a binary string, thereby enabling the computation of its complexity via Shannon entropy. Empirical analyses conducted on multiple real-world networks reveal an approximately linear relationship between the proposed entropy measure and the predictive performance of the best-performing link prediction algorithm for each dataset. 





In a related setting, \citet{lu2015toward} introduce an alternative metric for network predictability. This approach exploits the network's representation as a binary, symmetric adjacency matrix. Under the assumption that the structural consistency of networks subject to small perturbations can be approximated by the corresponding changes in the eigenvalues of the adjacency matrix, the authors perform an eigendecomposition of a perturbed version of this matrix (constructed by randomly removing a fraction of its links from the original adjacency matrix). They then quantify the structural discrepancy between the original and perturbed networks by evaluating the induced perturbation in the eigenvalue spectrum.


Neither of these recent studies can be applied directly to RS, whose user–item rating matrices are inherently non-square and weighted. Any method that claims to assess the structural complexity of RS must therefore be designed to explicitly handle these defining characteristics.
These methods are effective in scenarios where binary values dominate the analysis, a situation characterized by unweighted graph edges where the representation matrix is also symmetric. In this work, we aim to measure these properties on non-symmetric, non-square, non-binary matrices, which is the general case for collaborative filtering RS representations.

\subsection{Collaborative Filtering (CF) Algorithms}

CF algorithms commonly use traditional machine learning methods~\cite{aggarwal2016recommender}. Latent factor CF is a popular approach that represents users and items as vectors in a low-dimensional latent space \cite{koren2009matrix}. These latent factors are not directly observable but can be inferred from user–item ratings or interactions. Methods for learning this latent structure include non-negative matrix factorization~\cite{lee2000algorithms}, support vector machines~\cite{xia2006support}, probabilistic matrix factorization~\cite{wang2011collaborative}, and, more recently, deep neural networks~\cite{he2017neural}.

\subsection{Denoising in Recommendation}

The goal of denoising in RS is to identify informative interactions while suppressing those that contribute little or negatively to a model’s ability to capture generalizable user-item structure. 

Recent denoising approaches can be broadly categorized into two main groups. The first includes methods that perform sample drop or sample reweighting based on their estimated impact on a model’s training loss~\cite{he2024double,lin2023autodenoise,zhang2025shapley,ge2023automated,wang2023efficient}. Most of these approaches rely on training dynamics such as per-sample loss trajectories or predictive disagreement over time to infer whether interactions are noisy, and thus require model training to estimate interaction quality. As a result, they are inherently method-dependent and often computationally expensive.

The second category consists of methods that are largely agnostic to the underlying recommendation algorithm, relying instead on data or structure-level properties to identify unreliable interactions~\cite{wu2016collaborative,ye2023towards,tian2022learning}. For example, \citet{ye2023towards} propose a structure and embedding-level denoising framework that identifies unreliable interactions via neighborhood-based scores and through adversarial-style embedding perturbations.

Our work falls into this second category and complements existing structure-based approaches by introducing a global complexity-based perspective. Rather than relying on local structural cues, we quantify the contribution of interactions to the overall structural consistency and predictability of the user-item matrix.


\section{Structural complexity for RS}
\label{sec:metrics}

\subsection{Preliminaries}

RS aim is to infer which items individual users are most likely to perceive as relevant, based on their historical interaction patterns. This inference task can be modeled as a link prediction problem on a highly sparse bipartite user-item graph, where the vast majority of entries are unobserved, as individual users interact with only a small subset of the available items.

Formally, a collaborative filtering RS operates on a set of $n$ users and a set of $m$ items, which are jointly represented by an interaction matrix $M \in \mathbb{R}^{n \times m}$. Each entry of $M$ encodes the rating induced by the interaction between the user indexed by the corresponding row and the item indexed by the corresponding column. In real-world scenarios, such interaction matrices are typically characterized by extreme sparsity: the vast majority of entries are unobserved, as individual users interact with only a small subset of the available items. The goal of this work is to derive a comprehensive complexity measure for the matrix $M$ that explicitly accounts for the limited observational information in these sparse interaction patterns.

We tackle this challenge by introducing a new complexity metric that ranks RS datasets by how hard they are to predict—from those with simple, highly regular user–item interactions to those with rich, tangled, and resistant-to-prediction structures. At the core of our method is the SVD of the user–item interaction matrix, which reveals the data’s intrinsic dimensionality and structural complexity: datasets that can be accurately captured in a low-dimensional space are deemed inherently easier to predict, while those that demand high-dimensional representations reflect more complex, challenging preference patterns.



\subsection{Structural Consistency (SC)}

The method proposed by \citet{lu2015toward} exploits the spectral properties of network adjacency matrices, using first-order perturbations of eigenvalues to predict the likelihood of missing links. Their approach leverages Structural Consistency (SC), which assesses how structural perturbations affect the spectral characteristics of a graph. However, their framework assumes symmetric, binary adjacency matrices—constraints that do not hold for RS, where the user-item interaction matrix $M$ is inherently non-symmetric (rectangular) and may contain continuous rating values rather than binary indicators.

To overcome these limitations, we extend their eigenvalue perturbation framework to accommodate the inherently asymmetric structure of recommendation matrices by generalizing the spectral analysis to the SVD of the interaction matrix $M$. This extension enables the application of SC principles to bipartite user–item graphs, while explicitly accounting for the distinctive statistical and structural properties of CF data.

Our methodology is grounded in the hypothesis that the predictability of a recommendation dataset can be quantitatively evaluated via controlled perturbations of the user–item interaction matrix. In particular, if the random permutation of a subset of entries in \(M\) induces only negligible variations in the spectrum of the matrix, this implies that the underlying interaction patterns exhibit a high degree of structural regularity and are therefore more amenable to accurate modeling by recommendation algorithms. Conversely, if such perturbations substantially change the spectrum, the original matrix has weak intrinsic structure, implying lower predictability and greater difficulty for algorithmic inference.


\subsection{Adjacency Matrix Perturbation via Diagonalization}

Consider an undirected network represented by an adjacency matrix \(A\). The procedure introduced by~\citet{lu2015toward} removes a fraction \(p\) of the edges, resulting in a perturbed adjacency matrix \(A^{P}\), such that \(A = A^{P}  + \Delta A\). Since \(A^{P}\) is symmetric, it can be diagonalized as follows:
\begin{equation}
    A^P = XKX^T,
    \label{eq:lu-diag}
\end{equation}
where $X$ is an orthogonal matrix containing the eigenvectors of $A^P$, and $K$ is the diagonal eigenvalue matrix. 

The structural perturbation of $A$ is defined as the linear approximation of \(A\) based on the eigenvectors of $A^P$, as follows:

\begin{equation}
    \tilde{A} =  A^P + X \Delta K X^T = X (K+\Delta K) X^T,
    \label{eq:lu-perturb}
\end{equation}
where $\Delta K$ is the change in the eigenvalues given the fixed eigenvectors, which is approximated as follows:
\begin{equation}
    \text{Diag}(\Delta K) \approx \text{Diag}(X^T \Delta A X).
    \label{eq:a_diagonal}
\end{equation}

Finally, $\tilde{A}$ is used as a prediction for $A$, with the fraction of corrected links predicted serving as the predictability metric.

\subsection{SC via SVD}

We now adapt this method for SVD-based structural perturbations in RS. 
Consider a user–item interaction matrix \(M \in \mathbb{R}^{n \times m}\) with the set of observed entries denoted by \(\Omega = \{(i,j) : M_{ij} \neq 0\}\). We construct a matrix \(M^P\) by applying two complementary perturbation mechanisms controlled by the parameters \(p \in (0,1)\) (total perturbation fraction) and \(\alpha \in [0,1] \) (value perturbation ratio).

\textbf{Value Perturbation:} A fraction $\alpha p$ of randomly selected entries from $\Omega$ have their rating values shuffled while preserving the original sparsity pattern. Let $\Omega_{\text{val}} \subset \Omega$ denote this subset, with cardinality $|\Omega_{\text{val}}| = \lfloor \alpha p |\Omega| \rfloor$. This procedure quantifies the model's sensitivity to perturbations in individual rating values, without altering the underlying interaction structure.

\textbf{Structural Perturbation:} A fraction $(1-\alpha) p$ of entries from $\Omega$ are moved from their original positions to randomly sampled positions outside $\Omega$ (zero entries). Let $\Omega_{\text{remove}} \subset \Omega \setminus \Omega_{\text{val}}$ denote the entries to be removed, with $|\Omega_{\text{remove}}| = \lfloor (1-\alpha) p |\Omega| \rfloor$, and $\Omega_{\text{add}} \subset \Omega^c$ denote the positions where these entries are relocated, with $|\Omega_{\text{add}}| = |\Omega_{\text{remove}}|$ and $\Omega^c = \{(i,j): M_{ij} = 0\}$. This alters the bipartite graph structure by changing which user-item pairs have interactions, modifying the underlying interaction structure.

\textbf{Time-Weighted Sampling:} To more accurately emulate real-world recommendation settings, in which recent user–item interactions typically carry greater informational value, we apply a temporally aware sampling strategy when selecting entries for perturbation. Specifically, for each entry $(i,j)$ associated with timestamp $t_{ij}$, the probability of being selected is given by:

\begin{equation}
    P((i,j)) \propto \frac{t_{ij} - t_{\min}}{t_{\max} - t_{\min} + \epsilon}
\end{equation}
where $\epsilon > 0$ prevents division by zero for datasets with uniform timestamps.

The resulting perturbed matrix $M^P$ can be expressed as:
\begin{equation}
    M = M^P + \Delta M
\end{equation}
where $\Delta M$ captures both types of perturbations applied to the original matrix.

The SVD of $M^P$ is given by:
\begin{equation}
    M^P = U \Sigma V^T.
    \label{eq:b_r}
\end{equation}
We introduce matrices $\Delta U$, $\Delta V$, and $\Delta \Sigma$ such that:
\begin{equation}
    M = (U + \Delta U)(\Sigma + \Delta \Sigma)(V + \Delta V)^T.
    \label{eq:b}
\end{equation}

To estimate $\Delta U$ and $\Delta V$, we employ a computationally tractable approximation that leverages the symmetry properties of Gramian matrices.
We define the structural perturbation $\tilde{M}$ as:
\begin{equation}
    \tilde{M} = M^P + U \Delta \Sigma V^T =  U \Sigma V^T +  U \Delta \Sigma V^T  = U(\Sigma + \Delta \Sigma) V^T.
    \label{eq:tilde-m}
\end{equation}

Note that $\tilde{M}$ is defined in a manner analogous to $M$, with the important distinction that the former does not incorporate the terms $\Delta U$ or $\Delta V$. Our goal is to obtain an accurate approximation of $\Delta \Sigma$, since the remaining quantities can be obtained from $M^P$ and its SVD. 

We cannot directly reuse the approximation strategy employed for $\Delta K$ in Equation (\ref{eq:a_diagonal}), as that method is valid only when the original matrix is symmetric. Instead, we exploit the symmetry that arises from the product of a matrix with its transpose. To this end, we left-multiply Equation (\ref{eq:b_r}) by its transpose, thereby forming the Gram matrix:

\begin{equation}
     \mathcal{M}^P := (M^P)^T M^P = V \Sigma^T U^T U \Sigma V^T = V (\Sigma^T \Sigma) V^T,
\end{equation}
where we use the orthogonality property $U^T U = I$. Since $V$ is also orthogonal and $\Sigma^T \Sigma$ is diagonal, this provides a diagonalization of $(M^P)^T M^P$.

Taking the symmetric matrix $\mathcal{M}^P$ as the perturbed matrix in Equation (\ref{eq:lu-diag}) and $\mathcal{M} := M^T M$ as the original matrix $A$, we approximate $\Delta (\Sigma^T \Sigma)$ by applying steps analogous to those used for the matrix $A$, thereby obtaining the counterpart of Equation~(\ref{eq:a_diagonal}):

\begin{equation}
    \text{Diag}(\Delta (\Sigma^T \Sigma)) \approx \text{Diag}(V^T \Delta \mathcal{M} V),
    \label{eq:bb_diagonal}
\end{equation}
where:
\begin{equation}
    \Delta (\Sigma^T \Sigma) = (\Sigma + \Delta \Sigma)^T(\Sigma + \Delta \Sigma) - \Sigma^T \Sigma,
    \label{eq:sigma_squared}
\end{equation}
\begin{equation}
    \Delta \mathcal{M} =\mathcal{M} - \mathcal{M}^P
\end{equation}

For each diagonal element $i$, we solve for $\Delta \Sigma_{ii}$:

\begin{equation}
    \Delta \Sigma_{ii} \approx \sqrt{(\Sigma^T \Sigma)_{ii} + \Delta (\Sigma^T \Sigma)_{ii}} - \Sigma_{ii}.
    \label{eq:delta-sigma-approx}
\end{equation}

Since $\Delta (\Sigma^T \Sigma)= \text{Diag}(V^T \Delta \mathcal{M} V$), Equation~(\ref{eq:delta-sigma-approx}) provides an efficient approximation of $\Delta \Sigma$.

Our structural complexity framework is grounded directly in classical perturbation SVD theory~\cite{stewart1991perturbation}. Whereas the conventional analysis characterizes how perturbations affect both singular values and singular vectors, our approach instead projects the perturbation onto the singular vector space of the perturbed matrix. This formulation enables the evaluation of structural properties in the post-perturbation regime, thereby quantifying the stability of the resulting low-rank representation.

\subsubsection{Evaluation Metrics}
Once $\tilde{M}$ is computed using Equation~(\ref{eq:tilde-m}), it serves as an analytical approximation to $M$. 
We assess SC through two complementary metrics computed on the perturbed entry set $\Omega_{\text{pert}} = \Omega_{\text{val}} \cup \Omega_{\text{remove}} \cup \Omega_{\text{add}}$:

\textbf{SC RMSE:} The root mean squared error between true and analytically predicted ratings on perturbed entries:
\begin{equation}
    \label{Eq:RMSE}
    \text{RMSE} = \sqrt{\frac{1}{|\Omega_{\text{pert}}|} \sum_{(i,j) \in \Omega_{\text{pert}}} (M_{ij} - \tilde{M}_{ij})^2}
\end{equation}

To enable cross-dataset comparison, we normalize against baseline SVD reconstruction error:
\begin{equation}
    \label{Eq:SC_RMSE}
    \text{RMSE}_{\text{SC}} = \frac{\text{RMSE}}{\text{RMSE}_{\text{SVD}}}
\end{equation}
where $\text{RMSE}_{\text{SVD}}$ is the reconstruction error of standard SVD on the original matrix. Note that both metrics can be expressed using the Frobenius norm, namely, 
$\mathrm{RMSE} = 1\mathbin{/}\sqrt{|\Omega_{\text{pert}}|} \left\| \mathcal{P}_{\Omega_{\text{pert}}}( M - \tilde{M} ) \right\|_{F}$, where $\mathcal{P}_{\Omega_{\text{pert}}}(X)_{ij} = X_{ij}\, \text{, if } (i,j) \in \Omega_{\text{pert}}$, 0 otherwise. Therefore $\mathrm{RMSE}_{\mathrm{SVD}} = {1}\mathbin{/}{\sqrt{|\Omega_{\text{pert}}|}} \left\| \mathcal{P}_{\Omega_{\text{pert}}}( M - \tilde{M}_{\mathrm{SVD}} ) \right\|_{F}$ and $\mathrm{RMSE}_{\mathrm{SC}} = {\left\| \mathcal{P}_{\Omega_{\text{pert}}} ( M - \tilde{M} ) \right\|_{F}}\mathbin{/}\left\| \mathcal{P}_{\Omega_{\text{pert}}} ( M - \tilde{M}_{\mathrm{SVD}} ) \right\|_{F}$. We observe that the factor ${1}\mathbin{/}{\sqrt{|\Omega_{\text{pert}}|}}$ cancels out in $\mathrm{RMSE}_{\mathrm{SC}}$.

Lower $\text{RMSE}_{\text{SC}}$ values indicate that the dataset's structure is more resilient to perturbations, suggesting higher intrinsic predictability for recommendation algorithms.

\textbf{SC Spectral Distance:} The mean absolute change in singular values between the original and analytically reconstructed matrices:
\begin{equation}
    d_{\text{SC}} = \frac{1}{k} \sum_{i=1}^{k} |\sigma_i(M) - \sigma_i(\tilde{M})| = \frac{1}{k} \sum_{i=1}^{k} \Delta\Sigma_{ii}
\end{equation}
where $\sigma_i(\cdot)$ denotes the $i$-th singular value of a matrix and $k$ is the number of retained singular values.

\subsubsection{Algorithm Summary:} The complete structural perturbation method proceeds as follows:

\begin{enumerate}
\item \textbf{Matrix Perturbation:} Construct perturbed matrix $M^P$ by applying both perturbation types :
\begin{itemize}
\item $\Omega_{\text{val}}$: $\lfloor \alpha p |\Omega| \rfloor$ entries for value permutation.
\item $\Omega_{\text{remove}}$: $\lfloor (1-\alpha) p |\Omega| \rfloor$ entries for structural relocation.
\item $\Omega_{\text{add}}$: entries for relocated ratings.
\end{itemize}

\item \textbf{SVD Decomposition:} Compute rank-$k$ SVD of perturbed matrix: $M^P = U\Sigma V^T$.

\item \textbf{Perturbation Analysis:} Compute singular value corrections $\text{Diag}(\Delta(\Sigma^T\Sigma))$ using memory-efficient Gramian formulation.

\item \textbf{Analytical Reconstruction:} Derive corrected singular values $\tilde{\Sigma}_{ii} = \Sigma_{ii} + \Delta\Sigma_{ii}$ and reconstruct: $\tilde{M} = U\tilde{\Sigma}V^T$.

\item \textbf{Perturbation Error:} Compute metrics on $\Omega_{\text{pert}}$ using:
\begin{itemize}
\item $\text{RMSE}_{\text{SC}}$: accuracy of the perturbation method.
\item $\text{RMSE}_{\text{SVD}}$: Baseline SVD reconstruction accuracy.
\item $d_{\text{spec}}$: Relative spectral distance between true and approximated singular values.
\end{itemize}
\end{enumerate}

\subsubsection{Computational Optimization:} Computing $\text{Diag}(V^T \Delta \mathcal{M} V)$ requires materializing the full perturbation matrix $\Delta \mathcal{M} = \mathcal{M} - \mathcal{M}^P$, which has dimensions $m \times m$ and can exceed available memory for large datasets. Instead, we employ a memory-efficient reformulation that avoids explicit construction of $\Delta \mathcal{M}$.
Since $\mathcal{M} = M^T M$ and $\mathcal{M}^P = (M^P)^T M^P$, we can rewrite:
\begin{align}
\text{Diag}(V^T \Delta \mathcal{M} V) &= \text{Diag}(V^T (M^T M - (M^P)^T M^P) V) \\
&= \text{Diag}((MV)^T(MV) - (M^PV)^T(M^PV))
\end{align}

This reformulation computes only the diagonal elements through matrix-vector products $MV$ and $M^PV$ (each of size $n \times k$, where $k$ is the number of singular values), rather than forming the full $m \times m$ Gramian matrices. The computational complexity reduces from $O(m^2 k)$ to $O((n+m) \cdot \textit{nnz} \cdot k)$, where $\textit{nnz}$ is the number of non-zero entries in the sparse matrices, providing substantial memory savings for large-scale datasets.

\textbf{Truncated SVD Computation:} We further optimize by computing only partial SVD. Rather than computing the full SVD, we extract only the $k$ largest singular values and their corresponding singular vectors, where $k \ll \min(n,m)$. This approach leverages the low-rank structure of CF data and provides substantial computational savings by reducing SVD complexity.

\section{Experiments}
\label{sec:exps}

We conducted two complementary experiments to evaluate the proposed structural complexity metric. 
The first experiment investigates whether the metric adequately reflects \emph{empirical predictability}, operationalized as the attainable predictive accuracy of state-of-the-art RS algorithms on a given dataset. The second experiment evaluates the \emph{practical utility} of the metric by employing it to guide data selection for training RS models in data-scarce settings.
Collectively, these experiments are designed to test the following hypotheses:

\textbf{H1:} The proposed metric of structural complexity exhibits a systematic association with the empirically observed predictability of recommendation algorithms, such that increased structural complexity is linked to diminished recommendation performance.

\textbf{H2:} Structural complexity can be leveraged for data selection, thereby enabling RS to attain high performance while relying on comparatively smaller, yet higher-quality, datasets.




\subsection{Recommendation Algorithms}

Our experimental evaluation encompasses twenty-three recommendation algorithms that collectively represent the principal methodological paradigms in CF, chosen to reflect the progression from classical techniques to contemporary deep learning–based methods. As summarized in Table~\ref{tab:algorithms}, these methods are grouped into five categories: (i) traditional baselines, including popularity-based and neighborhood-based filtering; (ii) matrix factorization models that learn low-dimensional latent representations of users and items; (iii) neural network–based approaches that capture non-linear user–item interactions; (iv) graph neural networks that leverage the underlying graph structure of user–item interactions; and (v) generative or variational models that learn probabilistic representations of user behavior. All algorithms are implemented within the RecBole framework~\cite{zhao2021recbole}.

\begin{table}[t]
\centering
\caption{Selected methods and their references.}
\small
\begin{tabularx}{\columnwidth}{l X r}
\toprule
\textbf{Model} & \textbf{Family / Desc.} & \textbf{Ref.} \\
\midrule
\multicolumn{3}{l}{\textit{Traditional}} \\
Pop & Popularity-based & --- \\
ItemKNN & Item-based KNN & \cite{sarwar2001item} \\
\midrule
\multicolumn{3}{l}{\textit{Matrix Factorization}} \\
BPR & Bayesian Personalized Ranking & \cite{rendle2009bpr} \\
FISM & Factored Item Similarity Models & \cite{kabbur2013fism} \\
DMF & Deep Matrix Factorization & \cite{xue2017deep} \\
ENMF & Efficient Neural Matrix Factorization & \cite{chen2020efficient} \\
\midrule
\multicolumn{3}{l}{\textit{Neural Networks}} \\
NeuMF & Neural Collaborative Filtering (CF) & \cite{he2017neural} \\
NNCF & Neural CF with Interaction-based Neighborhood & \cite{bai2017neural} \\
ConvNCF & Outer Product-based Neural CF & \cite{he2018outer} \\
CDAE & Denoising Auto-Encoders & \cite{wu2016collaborative} \\
EASE & Embarrassingly Shallow Autoencoders & \cite{steck2019embarrassingly} \\
\midrule
\multicolumn{3}{l}{\textit{Graph Neural Networks}} \\
GCMC & Graph Convolutional Matrix Completion & \cite{berg2017graph} \\
SpectralCF & Spectral CF & \cite{zheng2018spectral} \\
NGCF & Neural Graph CF & \cite{wang2019neural} \\
DGCF & Disentangled Graph CF & \cite{wang2020disentangled} \\
LightGCN & Simplified GCN & \cite{he2020lightgcn} \\
SGL & Self-supervised Graph Learning & \cite{wu2021self} \\
SimpleX & Simple and Strong Baseline for CF & \cite{mao2021simplex} \\
\midrule
\multicolumn{3}{l}{\textit{Generative / VAE}} \\
MultiDAE & Denoising Autoencoder for CF & \cite{liang2018variational} \\
MultiVAE & Variational Autoencoder & \cite{liang2018variational} \\
MacridVAE & Disentangled Variational Auto-Encoder & \cite{ma2019learning} \\
RecVAE & Improved VAE & \cite{shenbin2020recvae} \\
\midrule
\multicolumn{3}{l}{\textit{Others}} \\
NCE-PLRec & Noise Contrastive Estimation for Linear Models & \cite{wu2018noise} \\
\bottomrule
\end{tabularx}
\label{tab:algorithms}
\end{table}



\subsection{Datasets}
\label{sec:datasets}

The datasets employed in the experiments were sourced from the repository maintained by RecBole\footnote{\url{https://recbole.io/dataset_list.html}}. Table~\ref{tab:datasets_stats} summarizes the basic statistics of these datasets.

\begin{table}[h!]
\centering
\caption{Datasets used in the experiments.}
\setlength{\tabcolsep}{2.5pt} 
\begin{tabular}{l c c c c c c}
\toprule
\textbf{Dataset} & \textbf{Users} & \textbf{Items} & \textbf{Inter.} & \textbf{Density} & \textbf{Inter/User} & \textbf{Inter/Item} \\
\midrule
Amazon\_Books & 10.3M & 4.4M & 29M & .0001 & 2.83 & 6.55 \\
Amazon\_Digital\_Music & 101K & 71K & 129K & .0018 & 1.28 & 1.83 \\
Amazon\_Movies\_and\_TV & 6,5M & 748K & 17.1M & .0004 & 2.64 & 22.95 \\
Amazon\_Grocery\ & 7M & 603K & 14M & .0003 & 2.00 & 23.34 \\
BeerAdvocate & 33K & 66K & 1.6M & .0719 & 47.52 & 24.02 \\
DianPing & 543K & 243K & 4,4M & .0034 & 8.15 & 18.18 \\
GoodReads & 876K & 2.3M & 228.6M & .0111 & 260.97 & 96.86 \\
KDD2010 & 2K & 585K & 2.3M & .2127 & 1244.42 & 3.91 \\
ModCloth & 48K & 1.3K & 83K & .1253 & 1.73 & 60.08 \\
RateBeer & 29K & 111K & 2.9M & .0903 & 99.92 & 26.43 \\
RentTheRunway & 106K & 6K & 193K & .0312 & 1.82 & 32.91 \\
anime & 74K & 11K & 7.8M & .9490 & 106.29 & 697.66 \\
book-crossing & 105K & 341K & 1.1M & .0032 & 10.92 & 3.38 \\
douban & 739K & 28 & 2.1M & 10.2741 & 2.88 & 75894.86 \\
epinions & 116K & 41K & 188K & .0039 & 1.62 & 4.6 \\
food & 227K & 232K & 1.1M & .0022 & 5.00 & 4.9 \\
jester & 73K & 100 & 4.1M & 56.3 & 56.3 & 41363.6 \\
lastfm & 1.8K & 18K & 93K & .2783 & 49.1 & 5.3 \\
ml-1m & 6K & 3.7K & 1M & 4.5 & 165.6 & 269.9 \\
netflix & 480K & 18K & 100.4M & 1.2 & 209.3 & 5654.5 \\
steam & 2.5M & 14K & 3M & .008 & 1.2 & 210.9 \\
twitch-100k & 100k & 740K & 3M & .004 & 30.5 & 4.1 \\
yahoo-music & 1.9M & 98K & 115.5M & .06 & 59.3 & 1176.8 \\
yelp & 1.9M & 209K & 8M & .002 & 4.1 & 38.3 \\
\bottomrule
\end{tabular}
\label{tab:datasets_stats}
\end{table}

\begin{figure*}[t]
   \centering
   \begin{subfigure}[b]{0.24\textwidth}
       \centering
       \includegraphics[width=\textwidth]{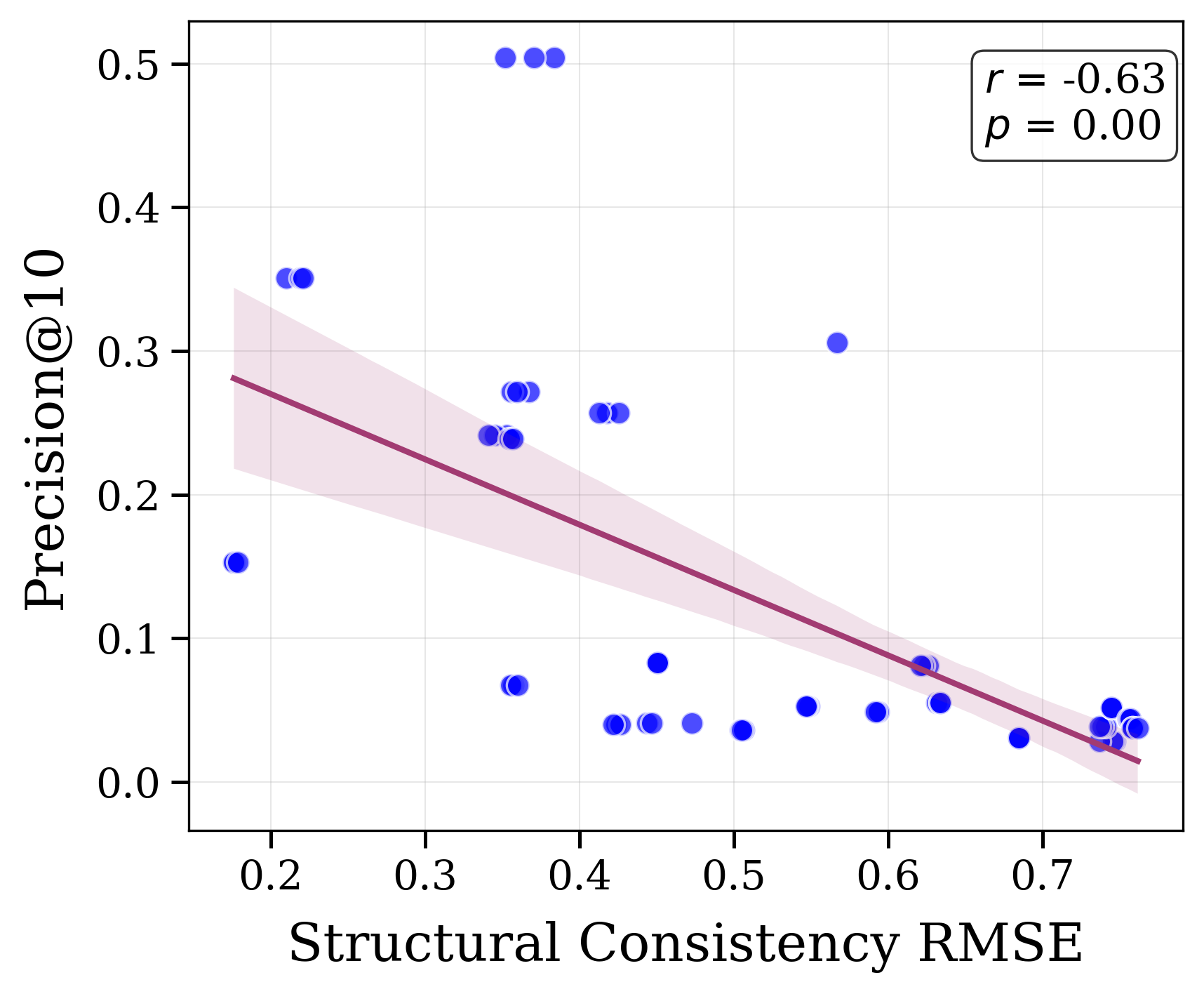}
       \caption{Precision@10}
   \end{subfigure}
   \begin{subfigure}[b]{0.24\textwidth}
       \centering
       \includegraphics[width=\textwidth]{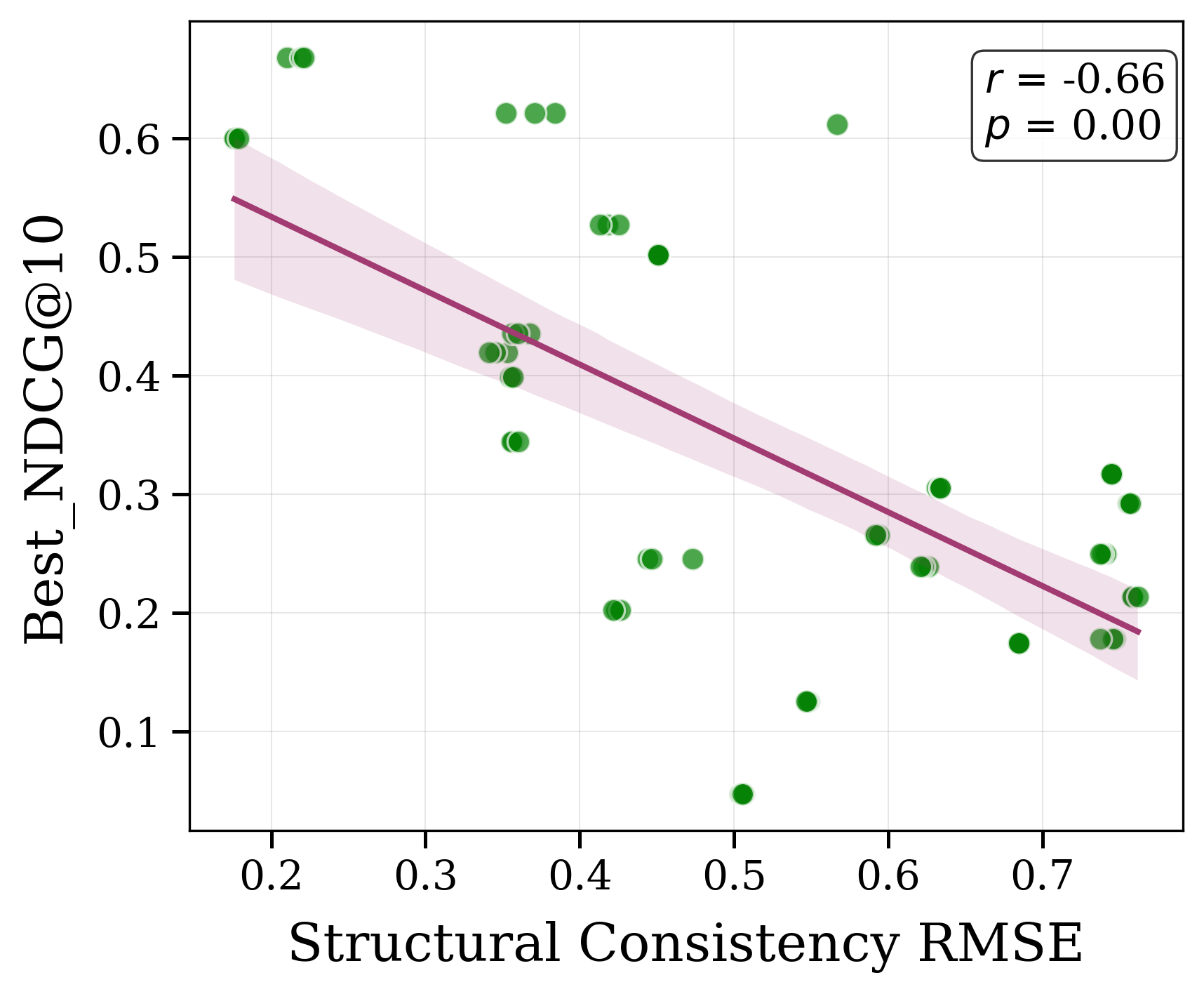}
       \caption{NDCG@10}
   \end{subfigure}
   \begin{subfigure}[b]{0.24\textwidth}
       \centering
       \includegraphics[width=\textwidth]{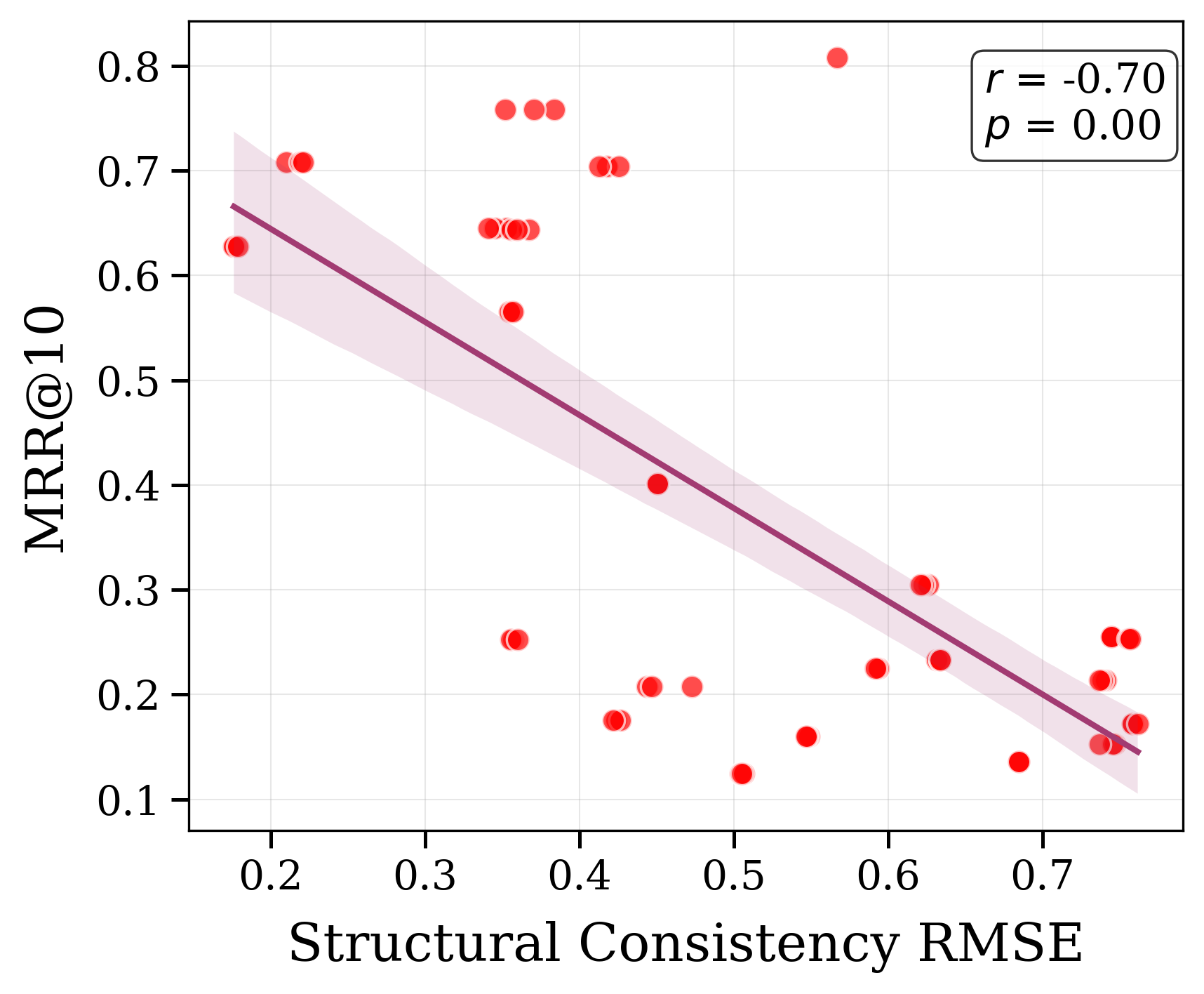}
       \caption{MRR@10}
   \end{subfigure}
   \begin{subfigure}[b]{0.24\textwidth}
       \centering
       \includegraphics[width=\textwidth]{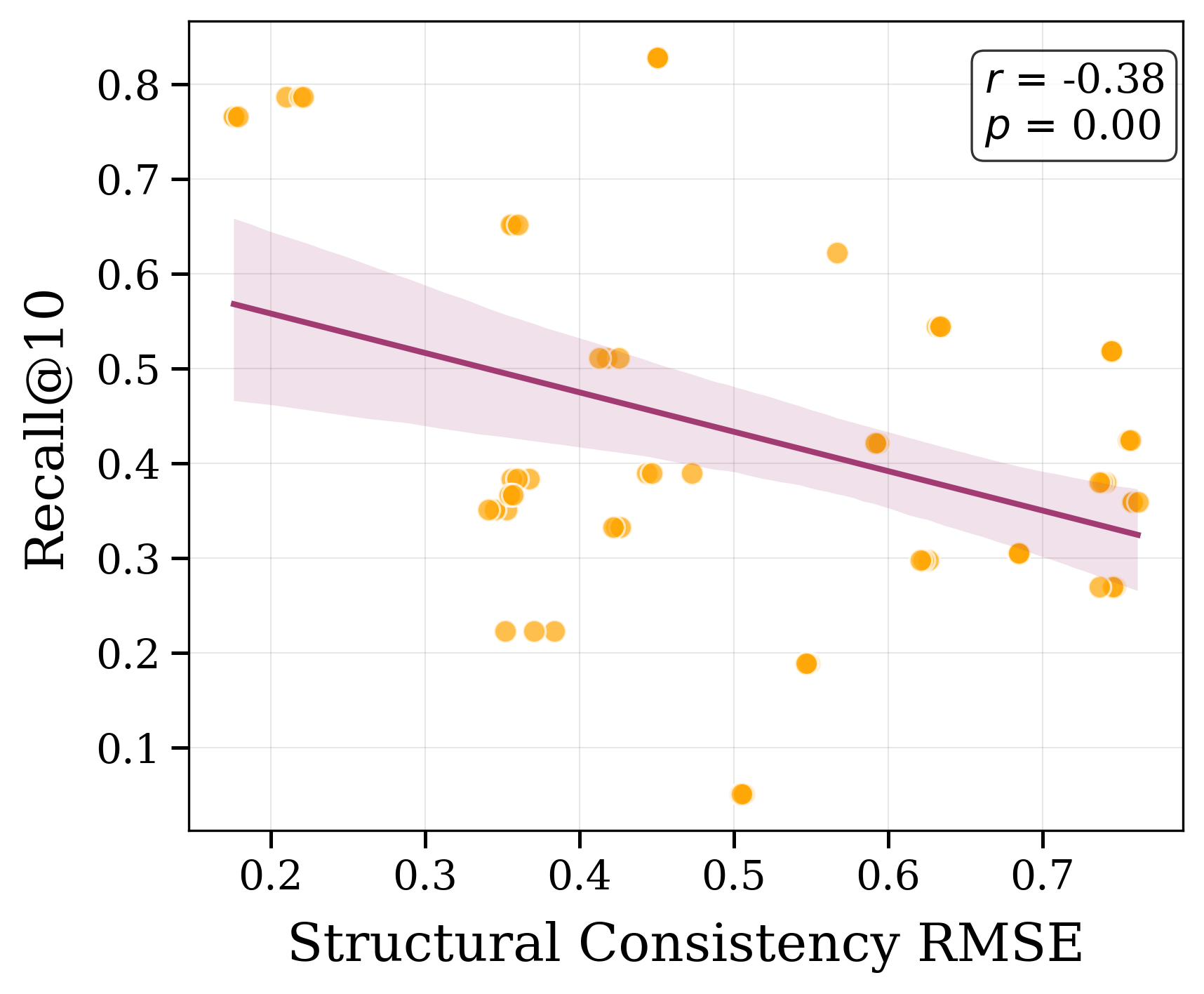}
       \caption{Recall@10}
   \end{subfigure}
   
   \caption{Correlation between $RMSE_{SC}$ and recommendation performance across metrics. Each point is a dataset with its best performance over all tested algorithms. Full results for all algorithms and datasets are in Appendix \ref{appendix:experiment_1} (Tables~\ref{tab:precision@10}, \ref{tab:recall@10}, \ref{tab:ndcg@10}, and \ref{tab:mrr@10}).}
   \label{fig:exp1_results}
\end{figure*}

\subsection{Experiment 1: Correlation with Empirical Predictability}

We empirically evaluate the extent to which our SC metrics capture the structural properties that govern the predictive performance of recommendation systems. Our working hypothesis is that datasets exhibiting high structural complexity—operationalized as the presence of patterns that are highly sensitive to perturbations—are intrinsically more challenging for recommendation algorithms to model and predict with high accuracy. In contrast, we posit that datasets with low structural complexity, characterized by clearer and more stable interaction patterns, enable recommendation algorithms to more effectively learn and exploit these regularities, thereby achieving superior predictive performance.

To empirically assess this hypothesis, we quantify the association between SC complexity scores and the highest attainable recommendation performance across a suite of state-of-the-art algorithms. For each dataset, we first compute the SC complexity score using our structural perturbation procedure, then train multiple recommendation methods and record the maximum performance achieved. This protocol is designed to ensure that the resulting correlation analysis reflects the intrinsic predictability of each dataset, rather than being confounded by the limitations or biases of any particular algorithm.



\subsubsection{Experimental Design}

To empirically evaluate our hypothesis that structural complexity serves as an estimator of recommendation system performance, we perform a correlation analysis designed to isolate the relationship between dataset complexity and algorithmic predictability. Specifically, for each dataset, we compute the best attainable performance across a suite of state-of-the-art recommendation algorithms, thereby ensuring that the resulting correlations predominantly capture intrinsic dataset properties rather than idiosyncratic limitations of individual algorithms.

\vspace{2mm}

\textbf{Data preprocessing}: We evaluate on 20 datasets spanning diverse recommendation domains, including movies and entertainment, e-commerce, food and restaurants, social media, books, beverages, and educational platforms.

Given that the recommendation datasets differ substantially in size, ranging from thousands to millions of interactions (see Table \ref{tab:datasets_stats}), we standardize each dataset to approximately $100{,}000$ interactions. This normalization facilitates a fair comparison of structural complexity measures and ensures computational tractability. 

To this end, we adopt a target-based sampling protocol inspired by \citet{mcelfresh2022reczilla}, The protocol begins by subsampling users from the original dataset in a density-aware manner to approximate a fixed interaction budget $N_{\text{target}} = 100{,}000$. Based on the interactions of the selected users, items are subsequently filtered to ensure sufficient support and maintain negative sampling feasibility.


For each dataset, we generate three independent samples using this procedure, resulting in a total of 60 dataset instances. Full methodological details and implementation specifics are provided in Appendix~\ref{appendix:sampling}.

\vspace{2mm}

\textbf{Performance Evaluation}: We evaluate recommendation quality using four complementary metrics: $Precision@10$, $NDCG@10$, $MRR@10$, and $Recall@10$. Together, these measures capture multiple dimensions of top-$k$ recommendation performance, including precision, recall, and ranking effectiveness. For each dataset instance, we record the maximum performance achieved across all twenty-three algorithms, thereby ensuring that our complexity–performance analyses are based on the best available algorithmic performance for each dataset.

All models are trained for a maximum of 10 epochs with early stopping applied when validation performance (MRR@10) fails to improve for 3 consecutive epochs. We use an 80/10/10 random split for the training, validation, and test sets. We also tested with an 80/10/10 temporal ordering split with similarly qualitative results (See the ablation studies in Section ~\ref{app:parameter_sensitivity} of the Appendix.).


\vspace{2mm}

\textbf{Structural Complexity Parameters:} For all structural complexity computations, we fix the perturbation fraction to $p = 0.1$ and the value perturbation ratio to $\alpha = 0.7$. Under this configuration, value shuffling is applied to $7\%$ of the interactions ($\alpha p$), while structural relocation is applied to $3\%$ of the interactions ($(1 - \alpha)p$). For the truncated singular value decomposition (SVD), we retain $k = 50$ leading singular values. A sensitivity analysis with respect to these hyperparameters is provided in Appendix~\ref{app:parameter_sensitivity}.




\subsubsection{Results}

Figure~\ref{fig:exp1_results} reports the correlation analysis between our normalized indicator of structural perturbation, $RMSE_{SC}$, and the maximal achievable performance of recommendation models across multiple evaluation metrics. Comprehensive, dataset-specific results are provided in Appendix~\ref{appendix:experiment_1} (see Table~\ref{tab:dataset_performance}). Each data point corresponds to a specific dataset, where the associated performance value denotes the maximum score achieved among all evaluated recommendation algorithms. 

Across all four metrics, we consistently observe a negative correlation between $RMSE_{SC}$ and the best method's performance on each dataset. This pattern indicates that performance tends to deteriorate as the internal structure of the interaction matrix becomes increasingly sensitive to perturbations, thereby providing empirical support for our hypothesis that datasets exhibiting higher structural complexity are intrinsically more challenging for recommendation algorithms.

The correlations are statistically significant across ranking-based metrics:

\begin{description}
    \item[--] \textbf{MRR@10} shows the strongest relationship ($r = -0.70$, $p < 0.001$), illustrating that structural complexity particularly impacts the ranking quality of top recommendations. 
    \item[--] \textbf{NDCG@10} exhibits a strong correlation ($r = -0.66$, $p < 0.001$), confirming that complex structural patterns hinder the algorithm's ability to properly rank relevant items.
    \item[--] \textbf{Precision@10} demonstrates a strong correlation ($r = -0.63$, $p < 0.001$), indicating that structural complexity significantly affects the precision of top-$k$ recommendations.
    \item[--] \textbf{Recall@10} shows a weaker correlation ($r = -0.38$), suggesting that while structural complexity impacts recall, this effect is less pronounced than for ranking-based metrics.
\end{description}

These correlation patterns exhibit strong robustness with respect to variations in hyperparameter configurations and train–test partitioning schemes, as evidenced by the sensitivity analysis reported in Appendix~\ref{app:parameter_sensitivity}.

The stronger correlations observed for ranking-based metrics (MRR@10, NDCG@10) relative to Recall@10 indicate that structural complexity primarily impairs the algorithm’s capacity to discriminate among relevant items and assign them appropriate ranks, rather than its ability to detect relevant items per se.

\textbf{Spectral Distance Results}: 
The spectral distance metric ($d_{spec}$) exhibits positive correlations with recommendation performance: Precision@10 ($r = 0.84$), MRR@10 ($r = 0.70$), and NDCG@10 ($r = 0.58$), with only Recall@10 showing no correlation ($r = -0.05$) (for detailed results, see Appendix \ref{appendix:experiment_1}, Figure~\ref{fig:exp1_results_sd}). 

Building on Stewart’s perturbation theory~\cite{stewart1991perturbation}, datasets that display greater spectral sensitivity under our reverse-projection framework can be interpreted as containing richer \emph{discriminative structure} that learning algorithms may leverage. In contrast to reconstruction-oriented complexity measures, our spectral distance quantifies the responsiveness of the singular value spectrum to perturbations that are propagated through the perturbed coordinate system, thereby characterizing the informativeness of the underlying structure rather than the difficulty of approximation.

\subsection{Experiment 2: SC-Guided Sampling for Data-Efficient Training}

The SC metrics capture dataset complexity by measuring how well structural patterns withstand perturbations. Each rating’s contribution to the overall $RMSE_{SC}$ shows how closely it follows the dataset’s structure: low perturbation errors indicate ratings well-predicted by the low-rank structure, while high errors indicate outliers or unpredictable interactions that deviate from global patterns.

This raises a key question: \emph{Do SC-based complexity scores effectively identify ratings that are most informative for training RS algorithms?} Clarifying this relationship is crucial to show that the proposed complexity metric is practically useful for algorithm training and optimization, not just for theoretical dataset characterization.

We hypothesize that ratings with low structural perturbation errors contain more essential structural information and are therefore more valuable for model training when data is limited. These "easy" ratings capture the core CF signals models must learn, while "difficult" ratings may reflect less generalizable ratings.

\subsubsection{Experimental Design}
We assess SC-guided sampling on a test set constructed by preserving each user’s final interaction, thereby inducing a temporally realistic evaluation scenario in which the models must forecast future user preferences. This test set is held fixed across all sampling strategies and sampling rates.

We compute per-rating structural complexity errors ($\text{RMSE}_{\text{SC}}$) by partitioning the full rating matrix into 10 disjoint subsets, each containing 10\% of all interactions. For each subset, we designate it as the perturbation set $\Omega_{\text{pert}}$, apply the structural perturbation procedure, and then compute prediction errors for the perturbed ratings. This yields an estimate of $\text{RMSE}_{\text{SC}}$ at the individual rating level.

We compare four sampling strategies:
\begin{itemize}
    \item \textbf{SC-Low Sampling} (primary): selects ratings with the lowest perturbation error.
    \item \textbf{SC-High Sampling} (control): selects ratings with the highest perturbation error.
    \item \textbf{Random Sampling} (baseline): uniformly samples ratings.
    \item \textbf{Temporal Sampling} (heuristic): selects the most recent ratings for each user.
\end{itemize}

From the available training interactions (i.e., all interactions excluding the fixed test set), we construct increasing-size subsets (10\%, 20\%, 30\%, \ldots, 100\%) for each sampling strategy. In all conditions, we employ per-user stratified sampling, ensuring that every user is represented in every subset; only the number of interactions per user varies with the subset size, while the set of users remains constant.

Recommendation models are trained on each sampled subset and evaluated on the fixed test set, facilitating a controlled comparison of how different sampling strategies preserve predictive performance as the volume of available training data is reduced.

\subsubsection{Results}

\begin{figure*}[ht]
   \centering
   \includegraphics[width=1.0\textwidth]{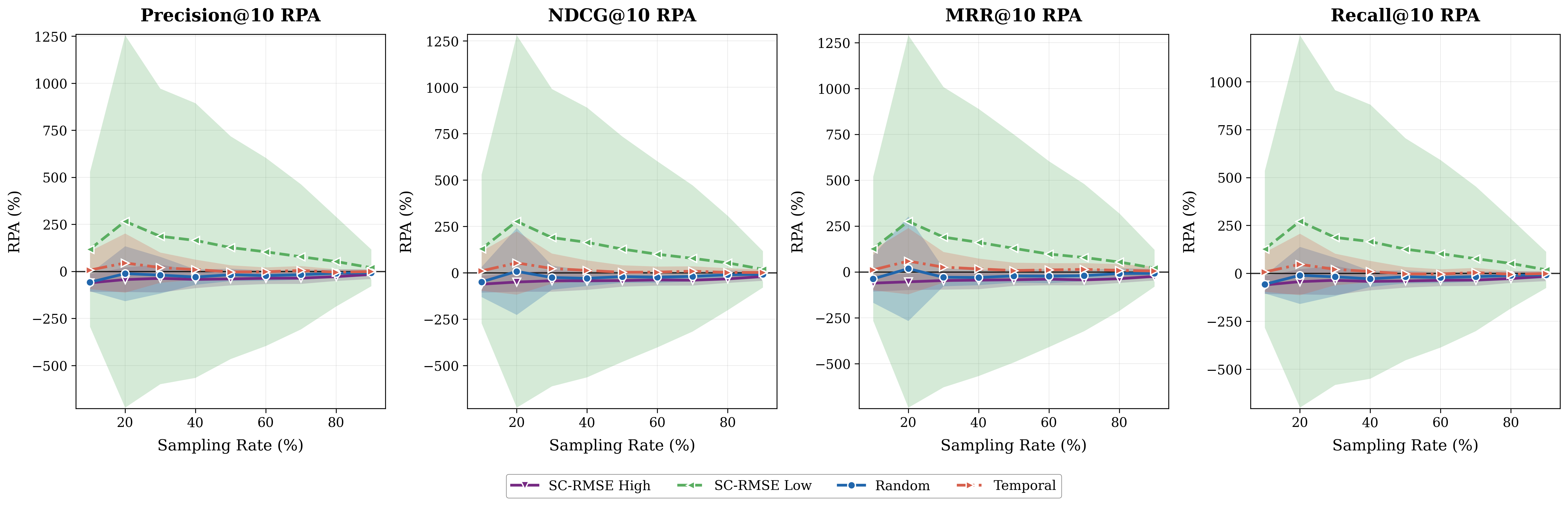}
      \caption{Relative Performance Analysis (RPA) illustrating the percentage change in LightGCN performance with respect to the 100\% sampling baseline under varying sampling strategies and sampling rates for the Precision@10, NDCG@10, MRR@10, and Recall@10 metrics. Shaded regions denote the standard deviation of the mean. Analogous trends are observed for the remaining methods (see Appendix \ref{appendix:cross_algorithm}, Figures~\ref{fig:bpr_rpa}, \ref{fig:ease_rpa}, and \ref{fig:MultiDAE_rpa}).}
   \label{fig:exp2_results}
\end{figure*}

Figure~\ref{fig:exp2_results} presents the Relative Performance Analysis (RPA) for SC-guided sampling strategies across different data availability scenarios. RPA measures the percentage change in performance relative to the 100\% baseline, defined as:
\[
\text{RPA} = \frac{\text{metric@X\%} - \text{metric@100\%}}{\text{metric@100\%}} \times 100,
\]
where positive values indicate improvement over training on the entire dataset and negative values indicate performance degradation.
The results support our central hypothesis that ratings with low structural perturbation errors contain more essential information for training recommendation algorithms.

\textbf{SC-Low Sampling} (green line) consistently outperforms all alternatives across metrics and sampling rates, with the largest gains under severe data scarcity. At 10\% sampling, SC-Low achieves approximately 250\% RPA across all evaluation metrics, indicating that a carefully selected 10\% subset can train models that perform 2.5 times better than those trained on the full dataset. This advantage persists across the sampling range, with gains gradually diminishing: approximately 150\% RPA at 20\%, 100\% at 30-40\%, and approaching baseline performance only at very high sampling rates (80-90\%).

However, substantial variance is observed in SC-Low sampling. This variance is most pronounced at low sampling rates—at 10\% sampling, the confidence interval spans from approximately -500\% to +1250\% RPA, indicating considerable variability in outcomes across different datasets. The variance gradually decreases as sampling rate increases. Despite this variance, SC-Low's mean performance consistently and substantially exceeds all baseline methods across all sampling rates and metrics.

\textbf{Temporal sampling} (orange line) maintains near-zero RPA values across most sampling rates, with modest positive gains (approximately 20-50\% RPA) only at very low sampling rates (10-20\%). This indicates that recency provides limited predictive value for training data selection, substantially less than SC-based approaches. The confidence intervals for temporal sampling are notably tighter than SC-Low, suggesting more consistent but modest performance.

\textbf{Baseline strategies}, random sampling (blue line) and SC-High sampling (purple line), show predominantly near-zero or slightly negative RPA values across all sampling rates. SC-High performs consistently at or below random selection, confirming that ratings with high perturbation errors are indeed less valuable for training.

To assess the generalizability of our findings, we replicated the analysis across representative algorithms from distinct methodological paradigms (Table~\ref{tab:algorithms}): BPR (Matrix Factorization),  EASE (Neural Networks), LightGCN (Graph Neural Networks), and MultiDAE (Generative/VAE). As shown in Appendix~\ref{appendix:cross_algorithm} (Figures~\ref{fig:bpr_rpa}, \ref{fig:ease_rpa}, and \ref{fig:MultiDAE_rpa}), SC-Low consistently achieves the highest mean RPA across all tested architectures. The magnitude of gains varies by algorithm: models like BPR and EASE show RPA values of 100-200\% at 10\% sampling, while neural architectures like MultiDAE exhibit even larger gains (300-500\%). This agreement across algorithms shows the robustness of structural complexity-based sampling as a general-purpose training strategy for RS.


\section{Conclusion}
\label{sec:conc}

Our results show that structural complexity is a powerful lens for understanding how predictable recommendation systems really are. Across datasets and metrics, higher structural complexity reliably coincides with poorer recommendation performance, providing strong support for hypothesis H1. This reveals that the intrinsic structure of user–item interaction graphs imposes hard limits on achievable predictive accuracy—limits that hold regardless of the modeling approach.

Beyond its explanatory value, structural complexity is directly useful for data selection. The standout performance of SC-Low sampling, especially under extreme data scarcity, supports hypothesis H2: a well-chosen subset of interactions can surpass training on the full dataset. 
This is evidence that interactions contribute unequally to learning, and that ratings with low structural perturbation error carry disproportionately rich training signals.

Overall, this work conceptualizes structural complexity not only as a diagnostic indicator of dataset complexity, but also as an operational instrument for intelligent, structure-aware data selection in RS training.

Several directions for future research remain open. Additional theoretical and empirical investigations are required to elucidate the two complementary notions of complexity introduced in this paper and to determine the contexts in which each metric is most appropriate, thereby enabling a more nuanced and rigorous characterization of datasets.
Extending structural complexity estimation to dynamic or streaming interaction graphs would enable online data selection and continual learning. Further work is needed to reduce computational overhead on very large datasets and to integrate structural-complexity signals directly into training objectives or active-learning pipelines. Finally, exploring applications beyond recommendations, such as link prediction and graph representation learning, could further establish structural complexity as a general tool for data-centric knowledge discovery.



\bibliographystyle{unsrtnat}
\bibliography{references}



\newpage

\appendix

\begin{figure*}[th]
   \centering
   \begin{subfigure}[b]{0.24\textwidth}
       \centering
       \includegraphics[width=\textwidth]{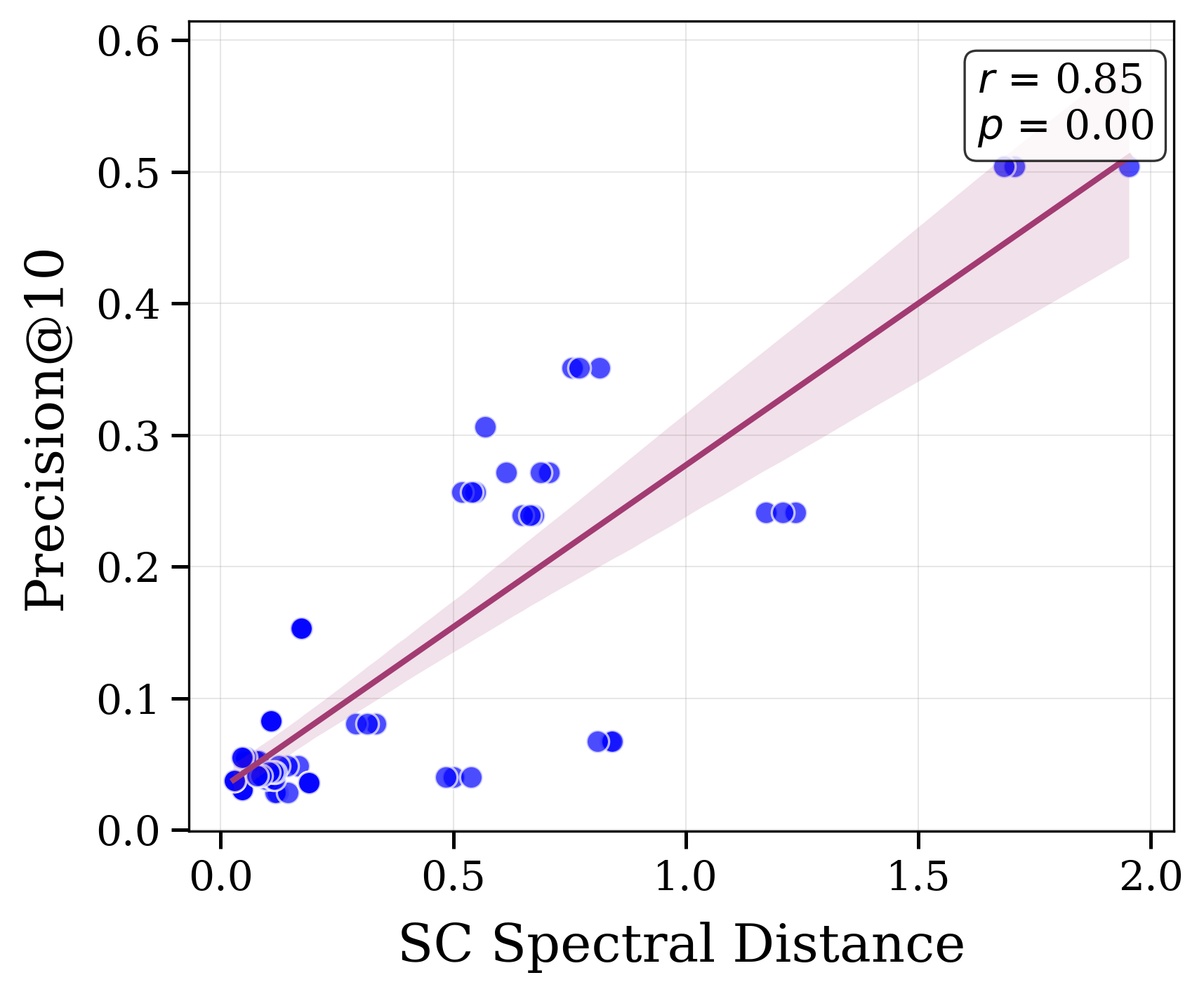}
       \caption{Precision@10}
   \end{subfigure}
   \begin{subfigure}[b]{0.24\textwidth}
       \centering
       \includegraphics[width=\textwidth]{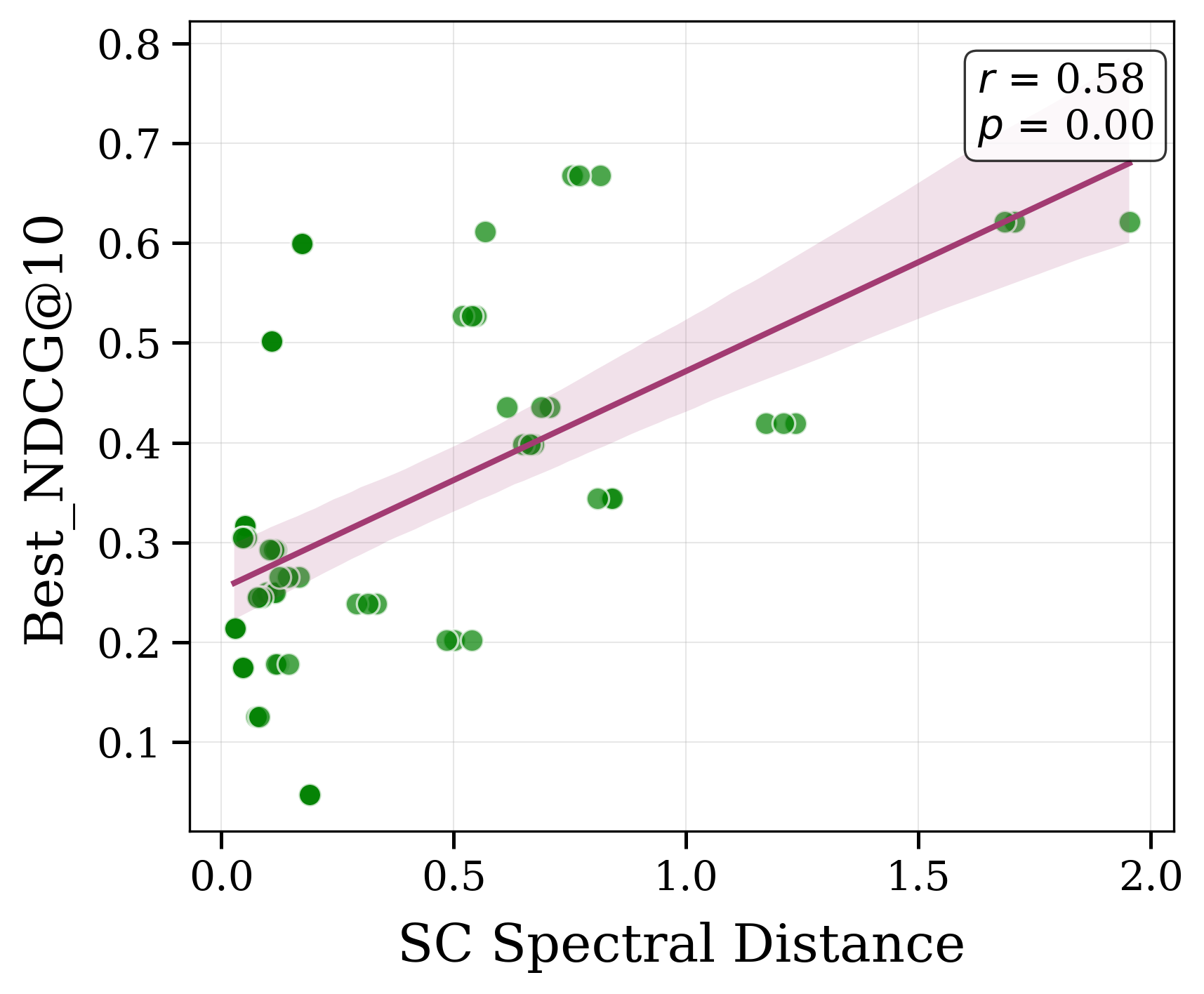}
       \caption{NDCG@10}
   \end{subfigure}
   \begin{subfigure}[b]{0.24\textwidth}
       \centering
       \includegraphics[width=\textwidth]{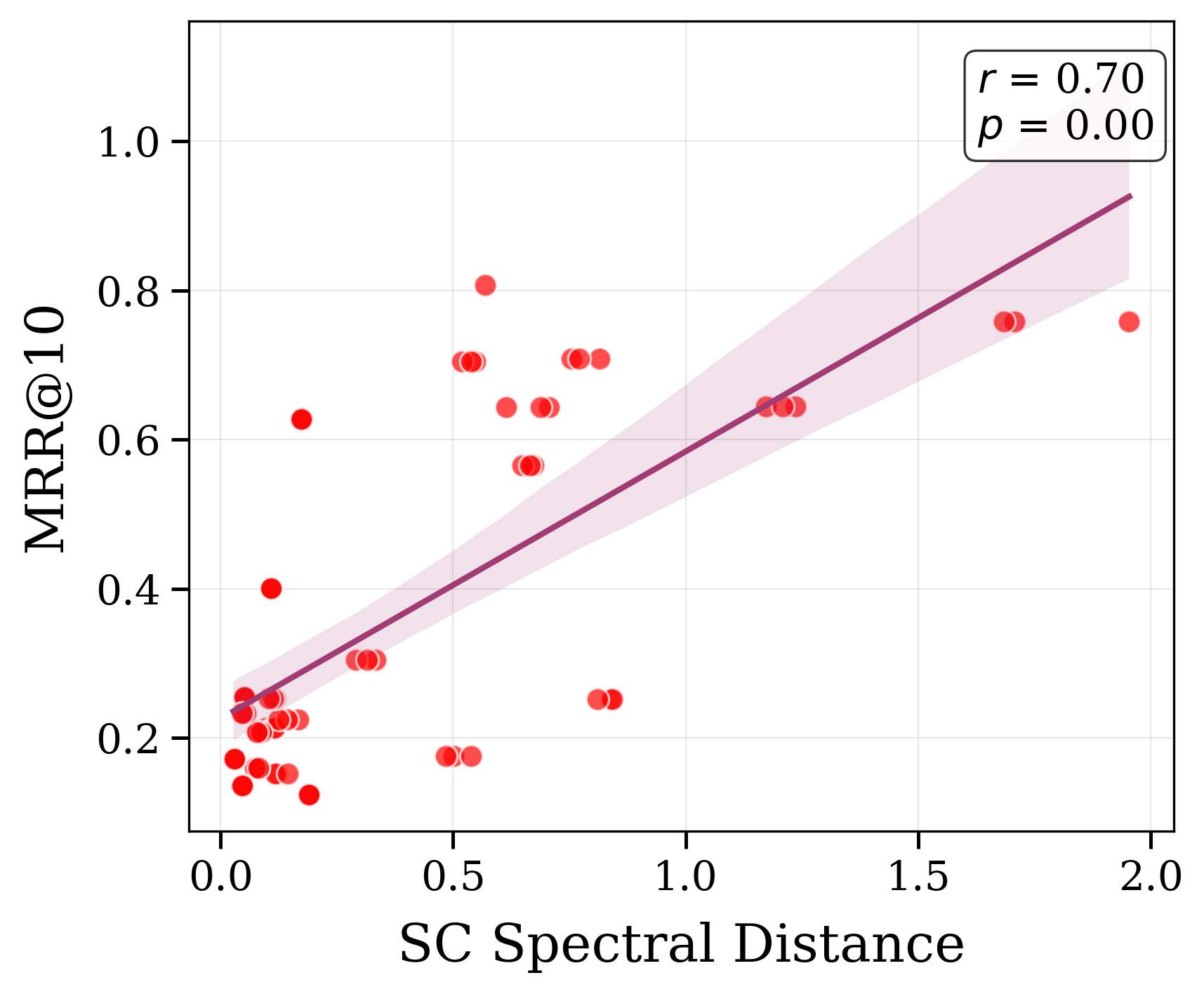}
       \caption{MRR@10}
   \end{subfigure}
   \begin{subfigure}[b]{0.24\textwidth}
       \centering
       \includegraphics[width=\textwidth]{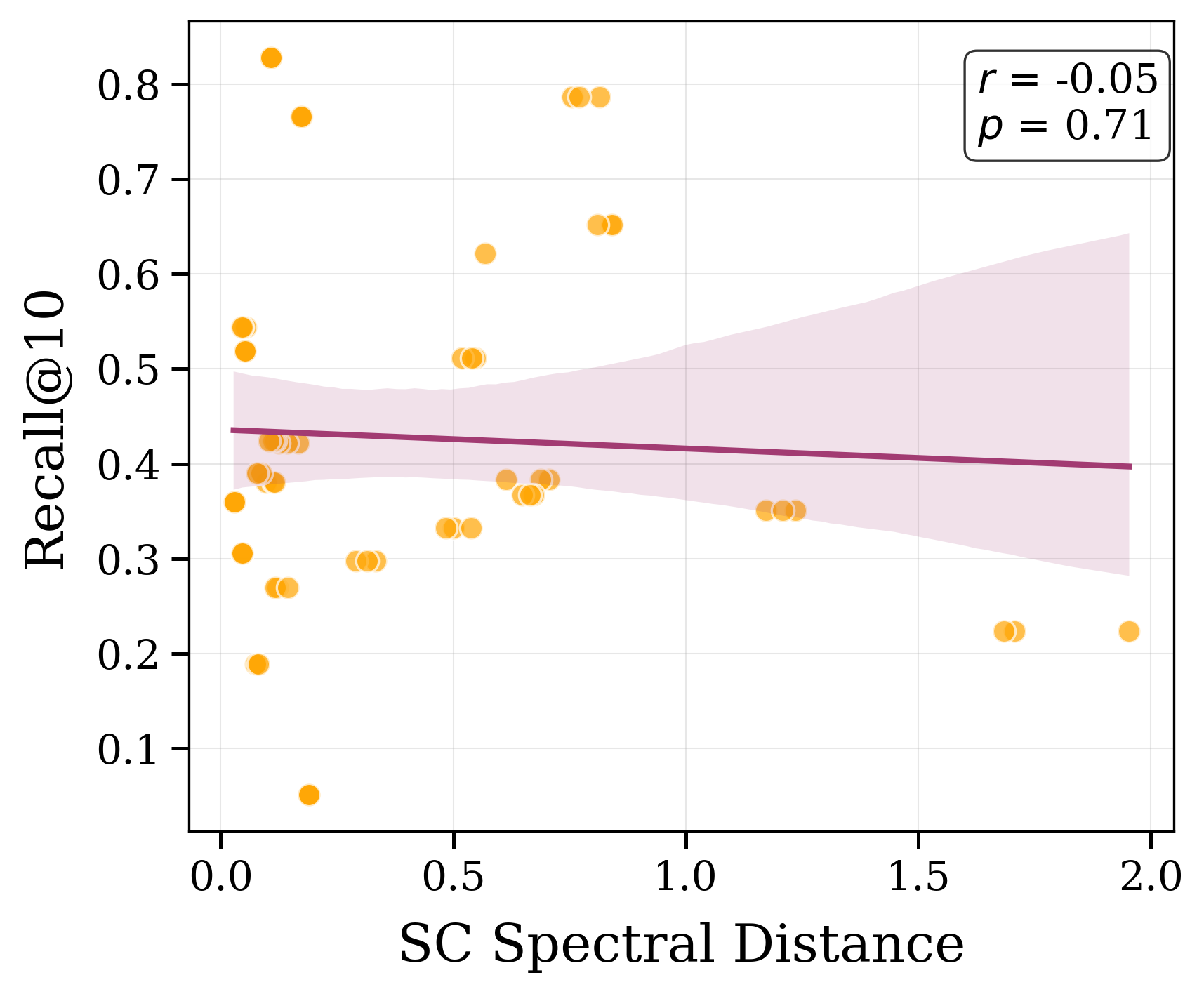}
       \caption{Recall@10}
   \end{subfigure}
   
   \caption{Correlation between the spectral distance ($d_{spec})$ and the best achievable recommendation performance across different evaluation metrics. Each point represents a dataset with its maximum performance achieved across all tested recommendation algorithms.}
   \label{fig:exp1_results_sd}
\end{figure*}

\section{Experiments}
\label{appendix:experiment}

\subsection{Sampling Details}
\label{appendix:sampling}

We evaluate the algorithms listed in Table~\ref{tab:algorithms} across multiple datasets using a sampling-based protocol. Due to the wide variation in dataset sizes, we normalize all datasets to a fixed interaction budget while preserving the structural properties. For each dataset, $n=3$ independent samples are generated, and all reported results correspond to averages computed across these samples.

Following the sampling principles presented by \citet{mcelfresh2022reczilla}, we employ a target-based procedure that constructs subsampled user-item interaction graphs with a fixed number of interactions, $N_{\text{target}} = 100{,}000$. The procedure consists of the following steps:

\paragraph{User Filtering and Subsampling.}
Users with fewer than five recorded interactions are excluded to mitigate cold-start effects. From the remaining users, we draw a random subsample. The target subsample size $N_{\text{users}}$ is determined by the interaction density of the original dataset, such that the expected number of interactions in the induced subgraph satisfies
\[
N_{\text{users}} \times \overline{I}_u \approx N_{\text{target}},
\]
where $\overline{I}_u$ denotes the average number of interactions per user in the original dataset.

\paragraph{Item Support Pruning.}
Given the induced user-item subgraph, items with fewer than two interactions are iteratively removed to ensure sufficient support for learning stable representations. Users left with no remaining interactions after this pruning step are discarded.

\paragraph{Negative Sampling Feasibility Check.}
Aggressive subsampling may lead to \emph{user saturation}, where a user has interacted with every item available in the subsampled dataset, making negative sampling infeasible. Let $I_u$ denote the set of items interacted with by user $u$, and let $I_{\text{sub}}$ denote the set of items present in the subsampled dataset. We enforce the constraint $|I_u| < |I_{\text{sub}}|$ for all users $u$.
When this condition is violated, we inject interactions with previously unseen items drawn from the original dataset; if this is not possible, the saturated user is removed from the subsample.

\paragraph{Target Truncation.}
If the total number of interactions exceeds $N_{\text{target}}$, interactions are randomly removed to match the target budget, while prioritizing the retention of injected interactions required to maintain negative sampling feasibility.

\subsection{Experiment 1}
\label{appendix:experiment_1}

\begin{table}[ht]
\centering
\caption{Best Performance Metrics per Dataset}
\label{tab:dataset_performance}
\begin{tabular}{lccccc}
\toprule
Dataset & P@10 & NDCG@10 & MRR@10 & R@10 & RMSE \\
\midrule
Amazon Books & 0.0284 & 0.1782 & 0.1528 & 0.2695 & 0.0981 \\
Amazon Movies and TV & 0.0386 & 0.2496 & 0.2137 & 0.3803 & 0.0985 \\
BeerAdvocate & 0.0526 & 0.1251 & 0.1603 & 0.1891 & 0.0992 \\
DianPing & 0.0489 & 0.2652 & 0.2252 & 0.4219 & 0.0990 \\
GoodReads & 0.2414 & 0.4194 & 0.6448 & 0.3512 & 0.0965 \\
KDD2010 & 0.5041 & 0.6213 & 0.7585 & 0.2233 & 0.0938 \\
ModCloth & 0.0828 & 0.5017 & 0.4012 & 0.8277 & 0.0775 \\
RateBeer & 0.0361 & 0.0473 & 0.1244 & 0.0515 & 0.0995 \\
RentTheRunway & 0.0519 & 0.3168 & 0.2551 & 0.5188 & 0.0863 \\
amazon digital music & 0.0440 & 0.2923 & 0.2529 & 0.4242 & 0.0977 \\
anime & 0.2568 & 0.5271 & 0.7044 & 0.5115 & 0.0958 \\
book-crossing & 0.0402 & 0.2020 & 0.1754 & 0.3326 & 0.0991 \\
douban & 0.0673 & 0.3439 & 0.2523 & 0.6521 & 0.0990 \\
epinions & 0.0308 & 0.1745 & 0.1359 & 0.3054 & 0.0980 \\
food & 0.0377 & 0.2136 & 0.1724 & 0.3596 & 0.0982 \\
jester & 0.3511 & 0.6676 & 0.7084 & 0.7867 & 0.0978 \\
lastfm & 0.3061 & 0.6114 & 0.8077 & 0.6219 & 0.0910 \\
ml-1m & 0.2717 & 0.4353 & 0.6438 & 0.3838 & 0.0995 \\
netflix & 0.2389 & 0.3986 & 0.5651 & 0.3669 & 0.0995 \\
steam & 0.0411 & 0.2452 & 0.2077 & 0.3900 & 0.0969 \\
twitch-100k & 0.0808 & 0.2386 & 0.3051 & 0.2977 & 0.0982 \\
yahoo-music & 0.1531 & 0.5998 & 0.6278 & 0.7659 & 0.0939 \\
yelp & 0.0553 & 0.3048 & 0.2333 & 0.5442 & 0.0981 \\
\bottomrule
\end{tabular}
\end{table}

\subsubsection{Experimental Results}

Table~\ref{tab:dataset_performance} summarizes the top performance attained for each evaluation metric and dataset across the 23 algorithms considered in our study. Detailed results for all algorithms, datasets, and metrics are reported in Tables~\ref{tab:precision@10}, \ref{tab:recall@10}, \ref{tab:ndcg@10}, and \ref{tab:mrr@10}.

\textbf{Spectral Distance Results.}
Figure~\ref{fig:exp1_results_sd} presents the correlation analysis between our spectral distance metric ($d_{spec}$) and recommendation system performance across evaluation metrics. In striking contrast to the $RMSE_{SC}$ results, the spectral distance exhibits predominantly \emph{positive} correlations with recommendation performance.

\begin{itemize}
    \item \textbf{Precision@10} demonstrates the strongest relationship ($r = 0.85$), indicating that datasets with higher spectral sensitivity achieve substantially better precision in top-k recommendations.
    \item \textbf{MRR@10} shows a strong positive correlation ($r = 0.70$), suggesting that spectral responsiveness particularly benefits ranking quality.
    \item \textbf{NDCG@10} exhibits a moderate positive correlation ($r = 0.58$), confirming that spectral distance captures discriminative structural properties.
    \item \textbf{Recall@10} shows no correlation ($r = -0.05$), indicating that the spectral distance metric is primarily associated with ranking precision rather than item coverage
\end{itemize}

These counterintuitive positive correlations reveal a fundamental insight into the structural complexity of recommendation systems. While reconstruction complexity ($RMSE_{SC}$) measures how difficult it is to approximate the original matrix from perturbed data, spectral distance measures how \emph{responsive} the singular value spectrum is when perturbations are projected through the perturbed coordinate system. Drawing from Stewart's perturbation theory~\cite{stewart1991perturbation}, datasets exhibiting higher spectral sensitivity under our reverse projection approach actually contain more \emph{discriminative structure} that algorithms can exploit.

\subsubsection{Ablation Study}
\label{app:parameter_sensitivity}

Our structural complexity metric depends on two key parameters: the perturbation fraction $p$ (which controls the total proportion of interactions perturbed) and the value perturbation ratio $\alpha$ (which controls the balance between value shuffling and structural relocation). Our main experiments use $p = 0.1$ and $\alpha = 0.7$, which balances computational efficiency. To assess the robustness of our approach to these hyperparameter choices, we conducted a sensitivity analysis across different parameter configurations.

\begin{figure}[t]
    \centering
    \begin{subfigure}[b]{0.49\columnwidth}
        \centering
        \includegraphics[width=\columnwidth]{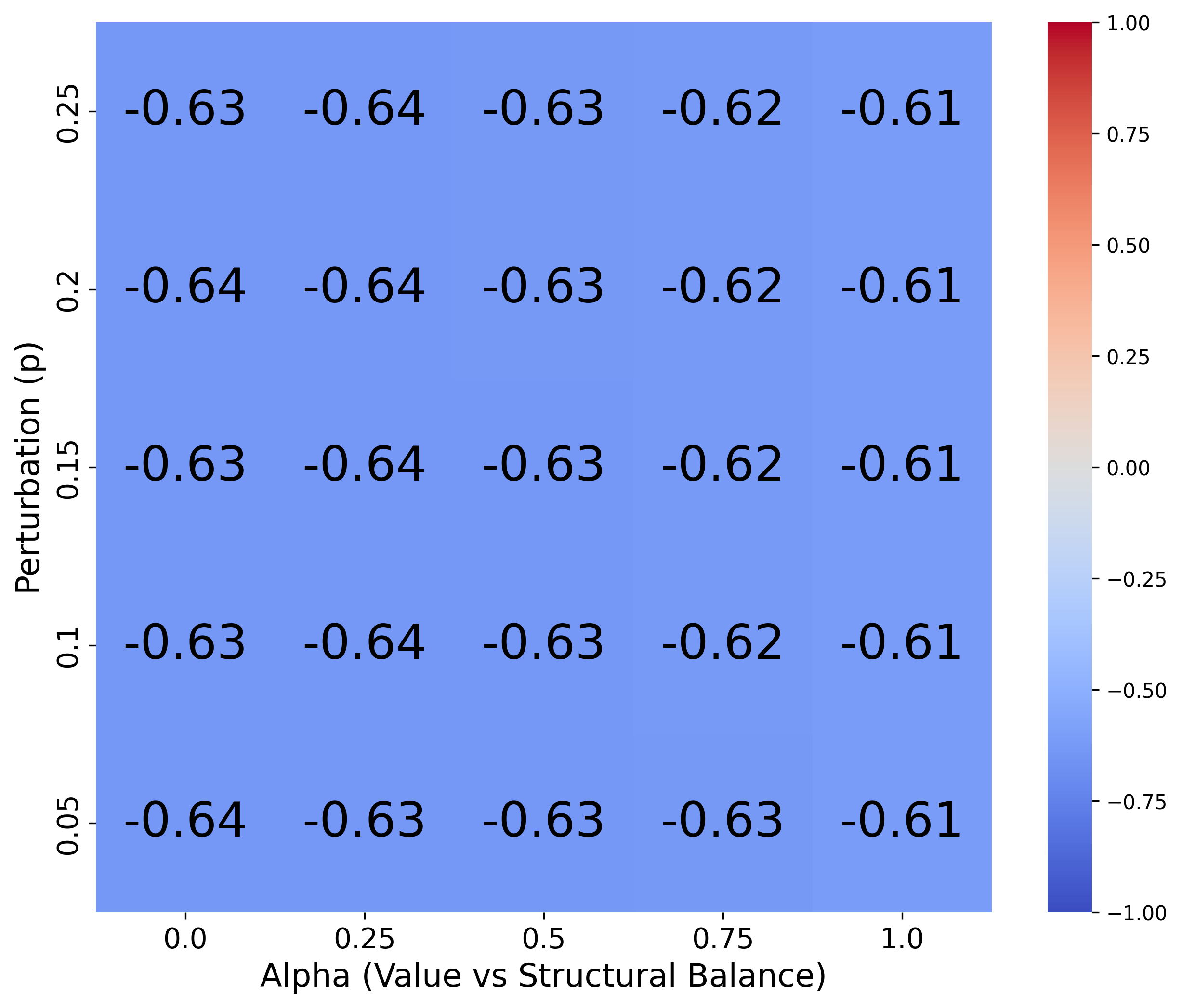}
        \caption{Precision@10}
        \label{fig:sens_precision}
    \end{subfigure}
    \begin{subfigure}[b]{0.49\columnwidth}
        \centering
        \includegraphics[width=\columnwidth]{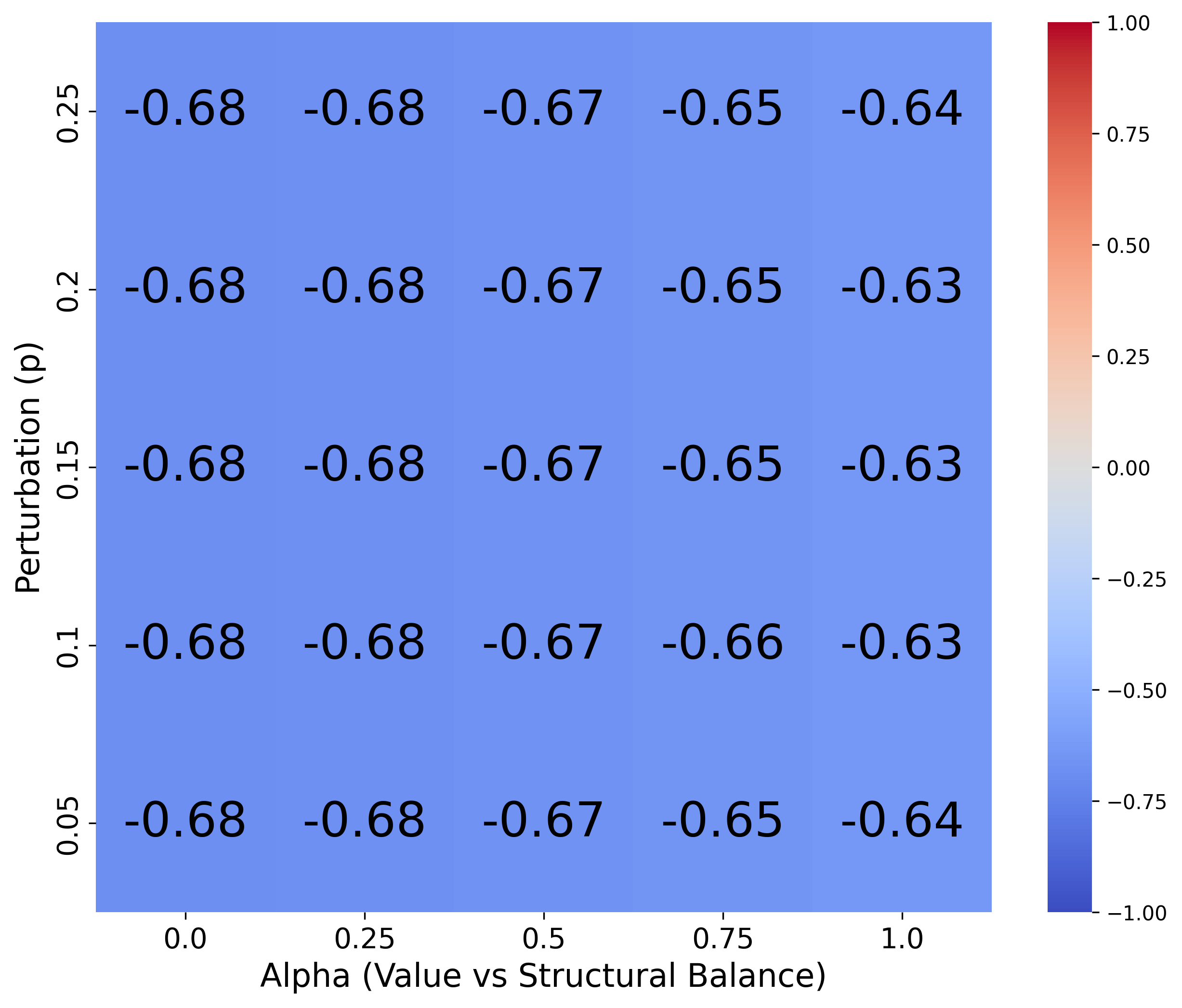}
        \caption{NDCG@10}
        \label{fig:sens_ndcg}
    \end{subfigure}
    
    \begin{subfigure}[b]{0.49\columnwidth}
        \centering
        \includegraphics[width=\columnwidth]{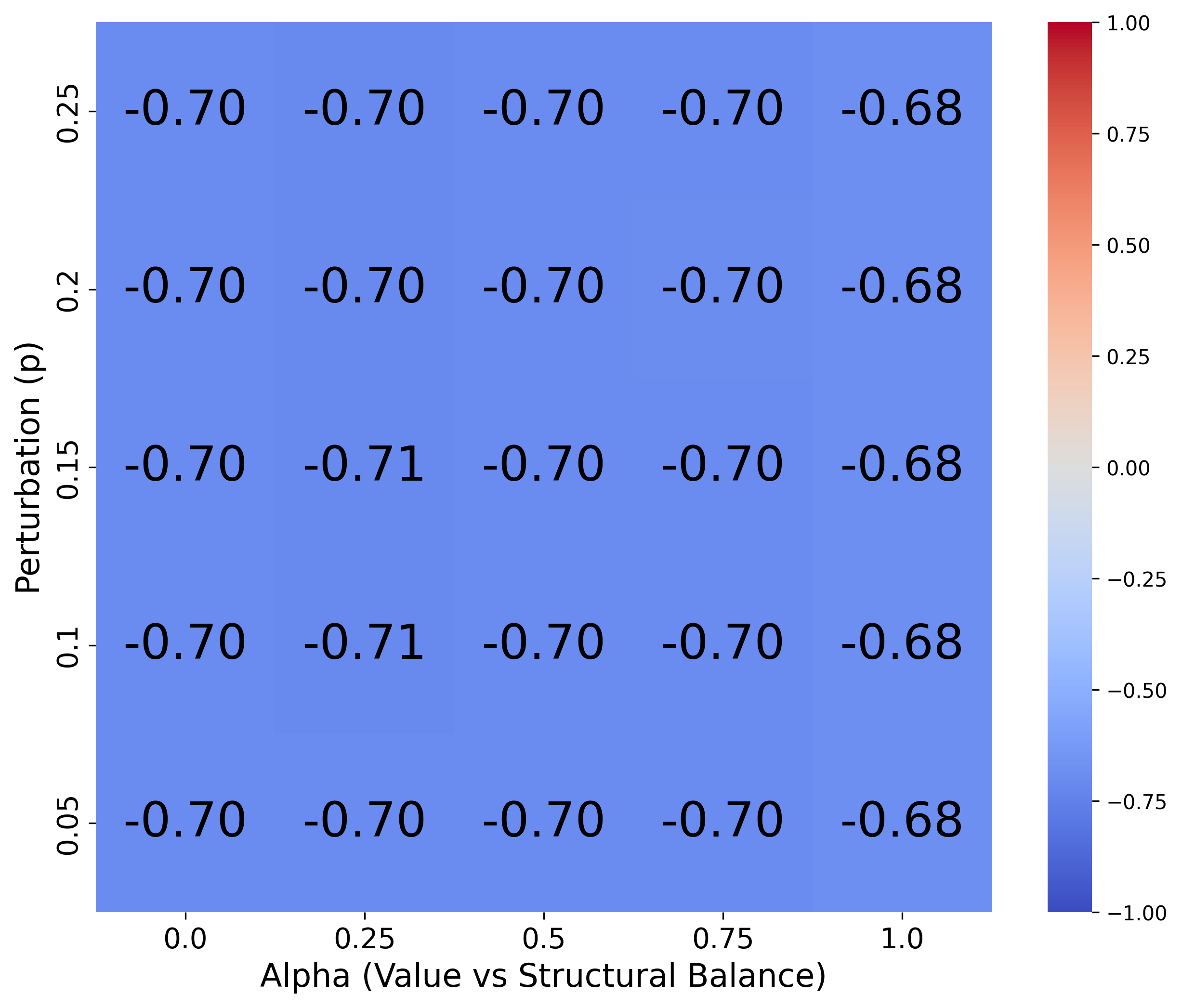}
        \caption{MRR@10}
        \label{fig:sens_mrr}
    \end{subfigure}
    \begin{subfigure}[b]{0.49\columnwidth}
        \centering
        \includegraphics[width=\columnwidth]{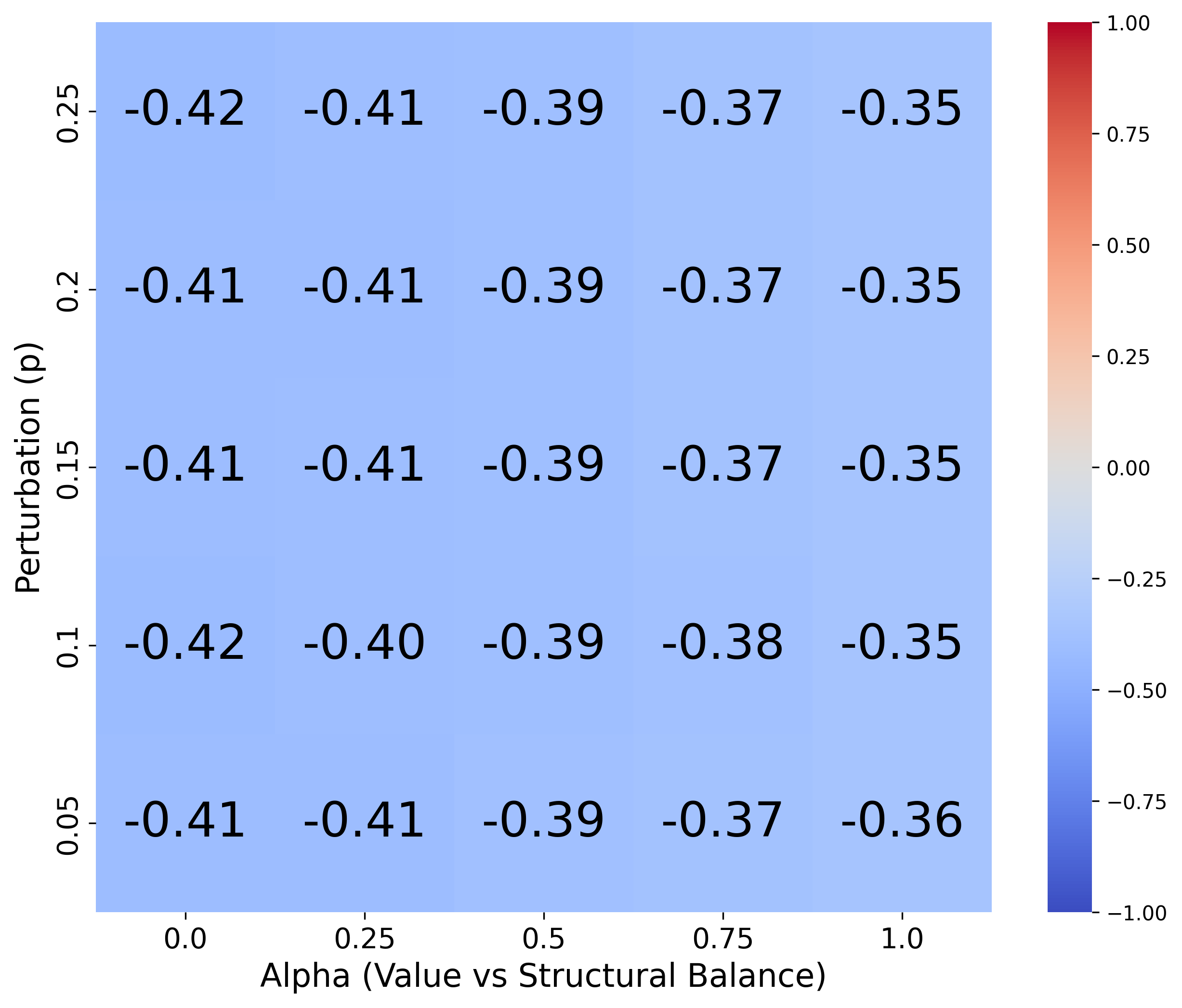}
        \caption{Recall@10}
        \label{fig:sens_recall}
    \end{subfigure}
    
    \caption{Sensitivity analysis of structural complexity metric to perturbation parameters $p$ and $\alpha$ under random train–test splitting. Each heatmap shows the Pearson correlation coefficient $RMSE_{SC}$ and the best achievable recommendation performance across all datasets. 
    }
    \label{fig:sensitivity_analysis}
\end{figure}

Figure~\ref{fig:sensitivity_analysis} shows the correlation coefficients between $RMSE_{SC}$ and the best recommendation performance across parameter configurations for four evaluation metrics. The results demonstrate stability, with variation of less than 0.07 across the entire parameter space. For ranking-based metrics (MRR@10, NDCG@10, Precision@10), correlations remain consistently strong (absolute values between $0.61-0.71$) regardless of the specific $(p, \alpha)$ configuration. This robustness confirms that our findings reflect properties of the dataset rather than artifacts of specific parameter choices.

\begin{figure}[t]
    \centering
    \begin{subfigure}[b]{0.49\columnwidth}
        \centering
        \includegraphics[width=\columnwidth]{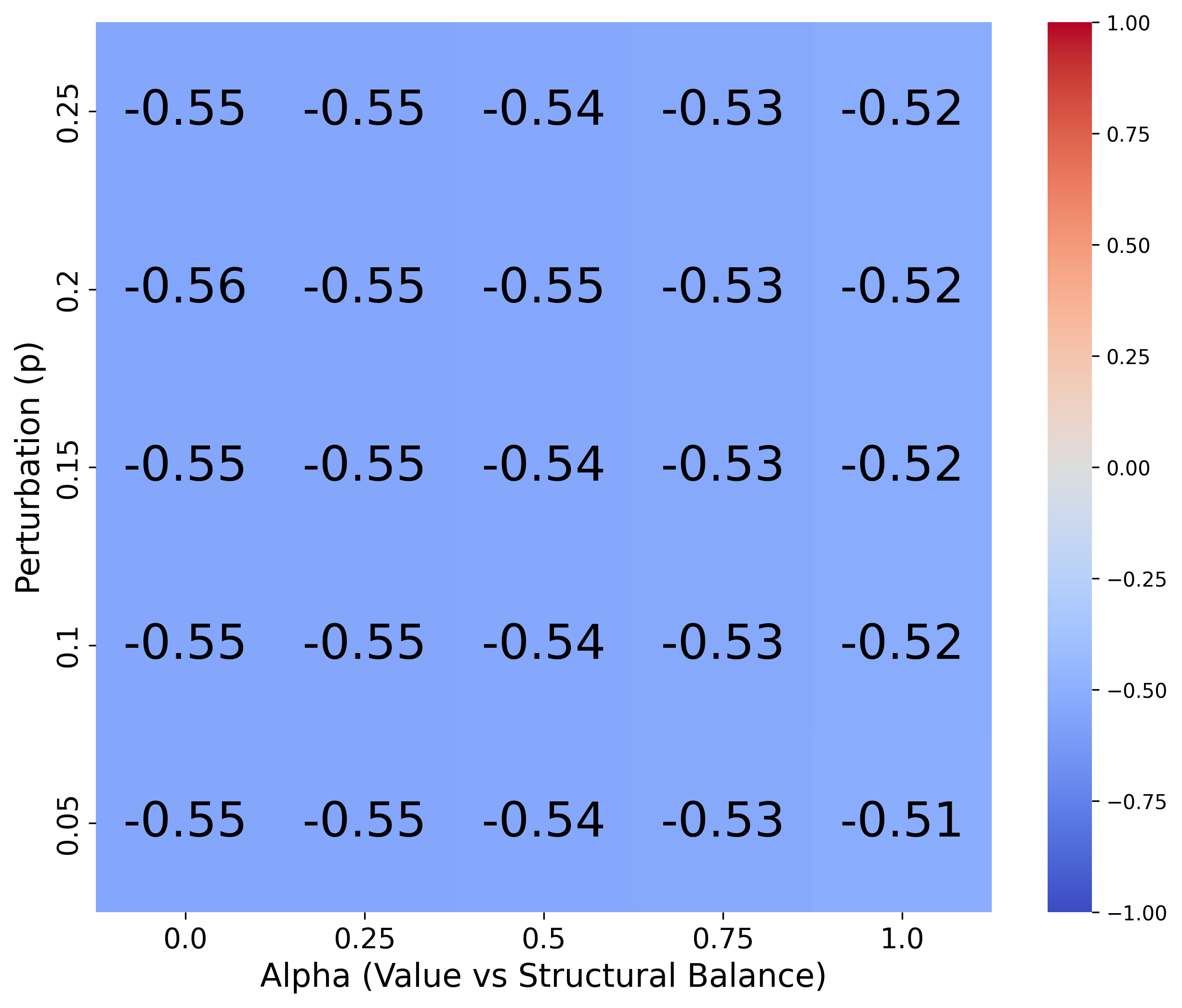}
        \caption{Precision@10}
        \label{fig:sens_precision}
    \end{subfigure}
    \begin{subfigure}[b]{0.49\columnwidth}
        \centering
        \includegraphics[width=\columnwidth]{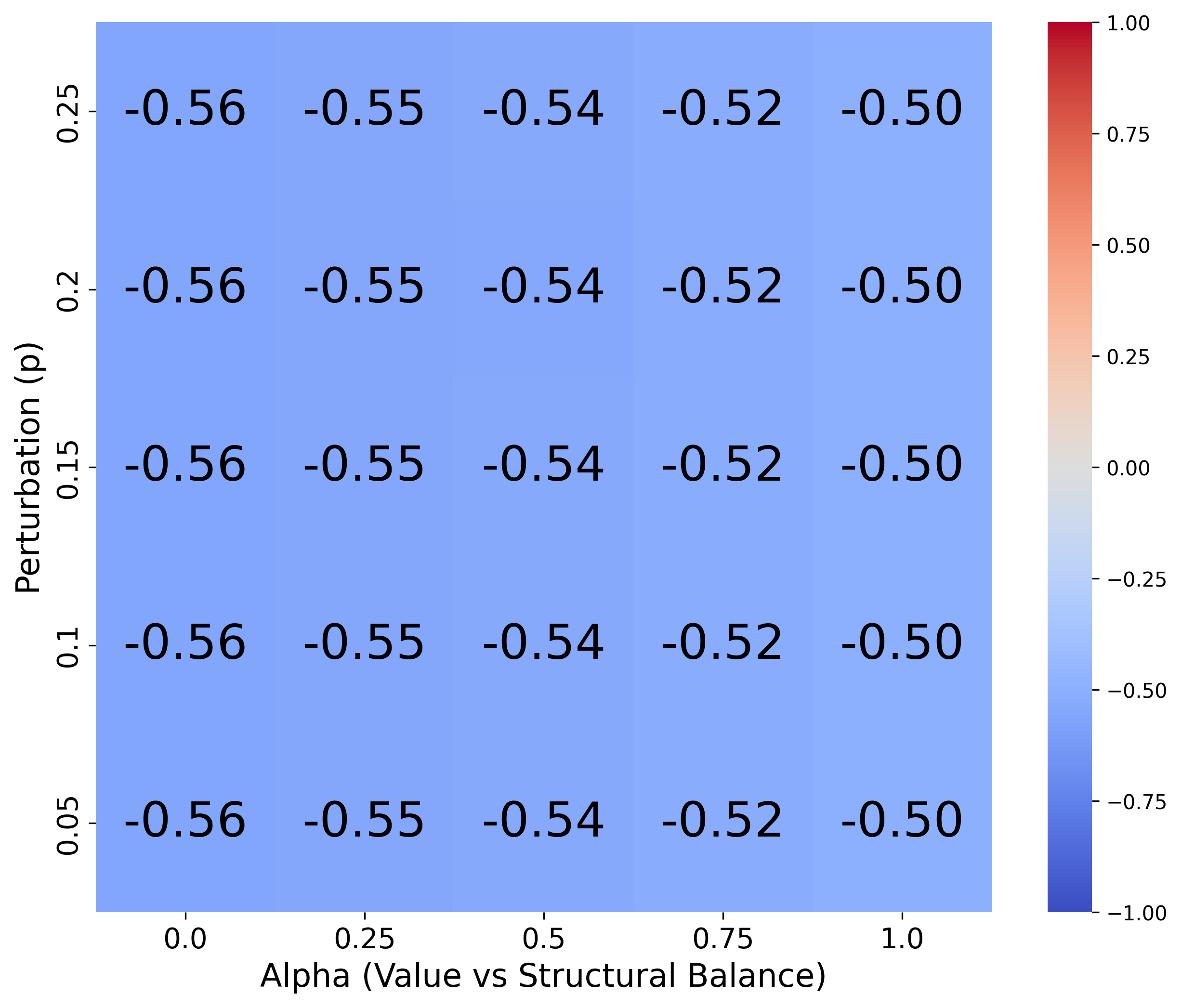}
        \caption{NDCG@10}
        \label{fig:sens_ndcg}
    \end{subfigure}
    
    \begin{subfigure}[b]{0.49\columnwidth}
        \centering
        \includegraphics[width=\columnwidth]{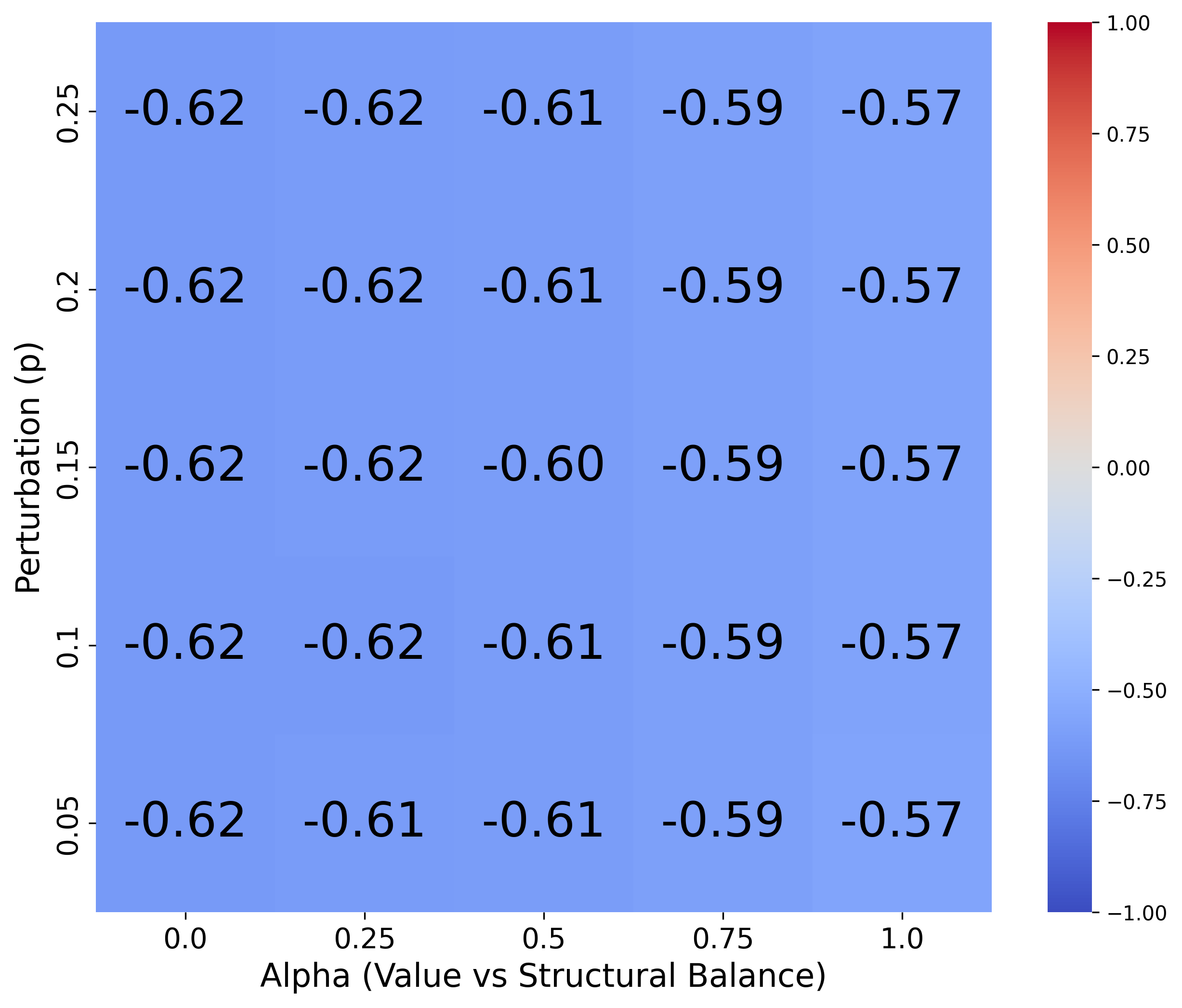}
        \caption{MRR@10}
        \label{fig:sens_mrr}
    \end{subfigure}
    \begin{subfigure}[b]{0.49\columnwidth}
        \centering
        \includegraphics[width=\columnwidth]{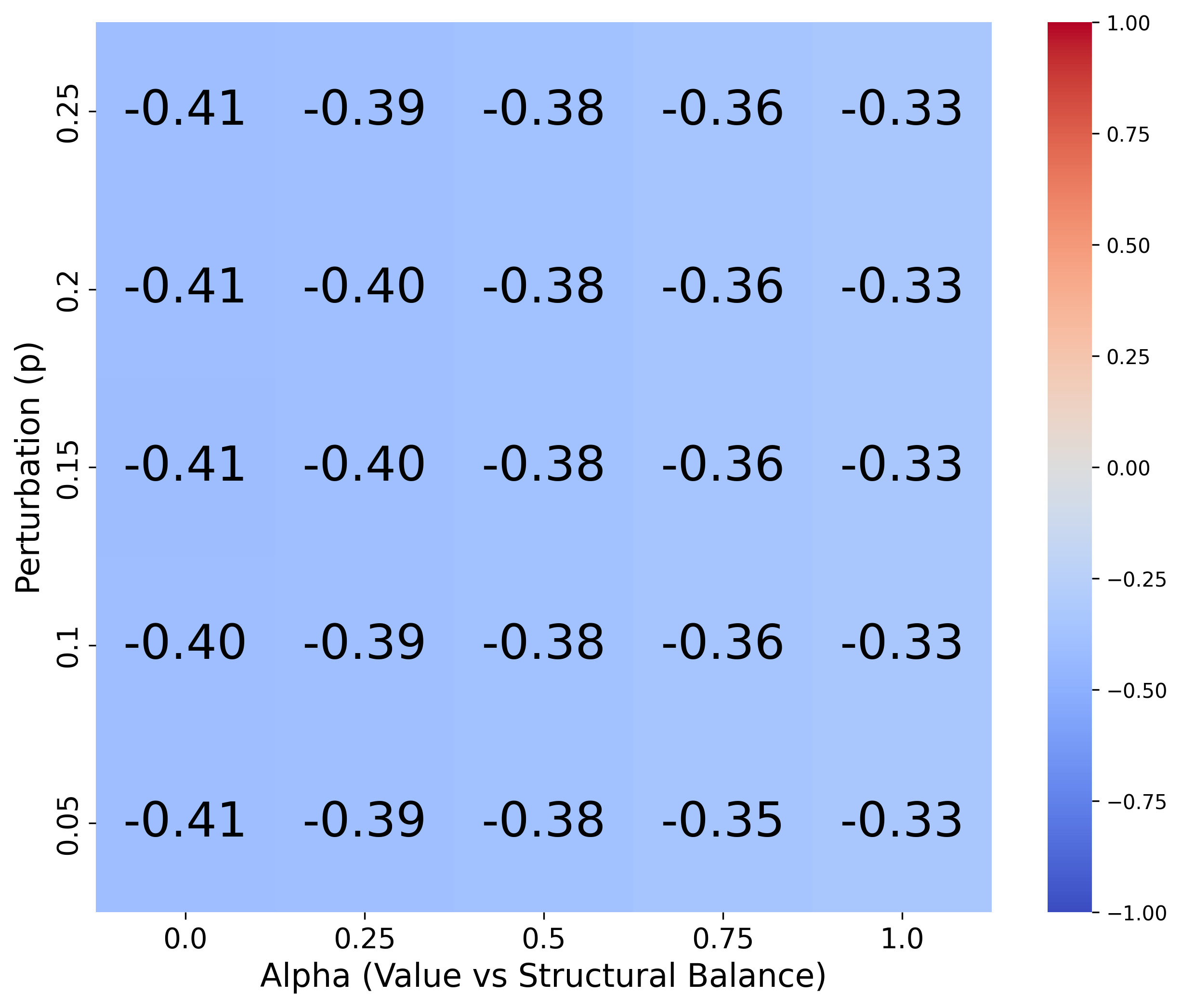}
        \caption{Recall@10}
        \label{fig:sens_recall}
    \end{subfigure}
    
    \caption{Sensitivity analysis of structural complexity metric to perturbation parameters $p$ and $\alpha$ under temporal train–test splitting. Each heatmap shows the  Pearson correlation coefficient $RMSE_{SC}$ and the best achievable recommendation performance across all datasets. 
    }
    \label{fig:sensitivity_analysis_TO}
\end{figure}

Additionally, we conducted a sensitivity analysis of the data-splitting strategy, comparing random ordering with temporal ordering on the test set. Figure~\ref{fig:sensitivity_analysis_TO} reports the correlation coefficients obtained under temporal splitting. Although the correlations are weaker than those observed with random splits, a clear moderate-to-high correlation signal persists, indicating that the relationship remains robust across splitting methods.

\subsection{Experiment 2: Cross-Algorithm Validation}
\label{appendix:cross_algorithm}

\begin{figure*}
   \centering
   \includegraphics[width=0.9\textwidth]{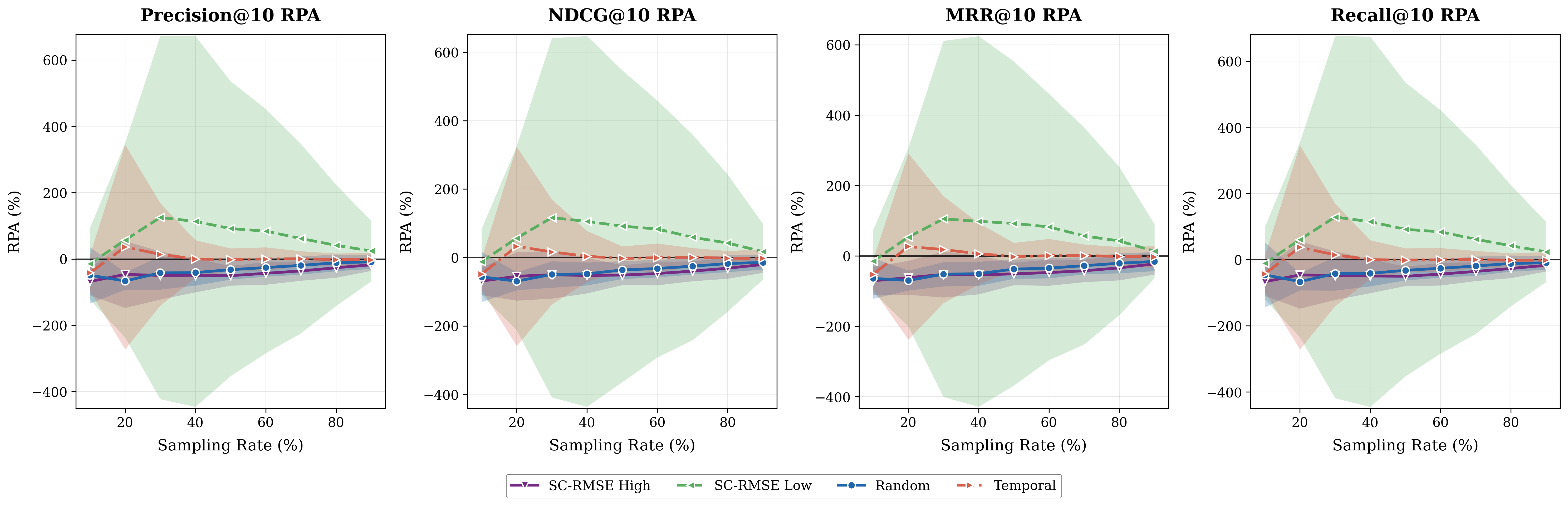}
   \caption{Relative Performance Analysis for BPR algorithm across different sampling strategies. Shaded areas represent the standard deviation of the mean.}
   \label{fig:bpr_rpa}
\end{figure*}

\begin{figure*}
   \centering
   \includegraphics[width=0.9\textwidth]{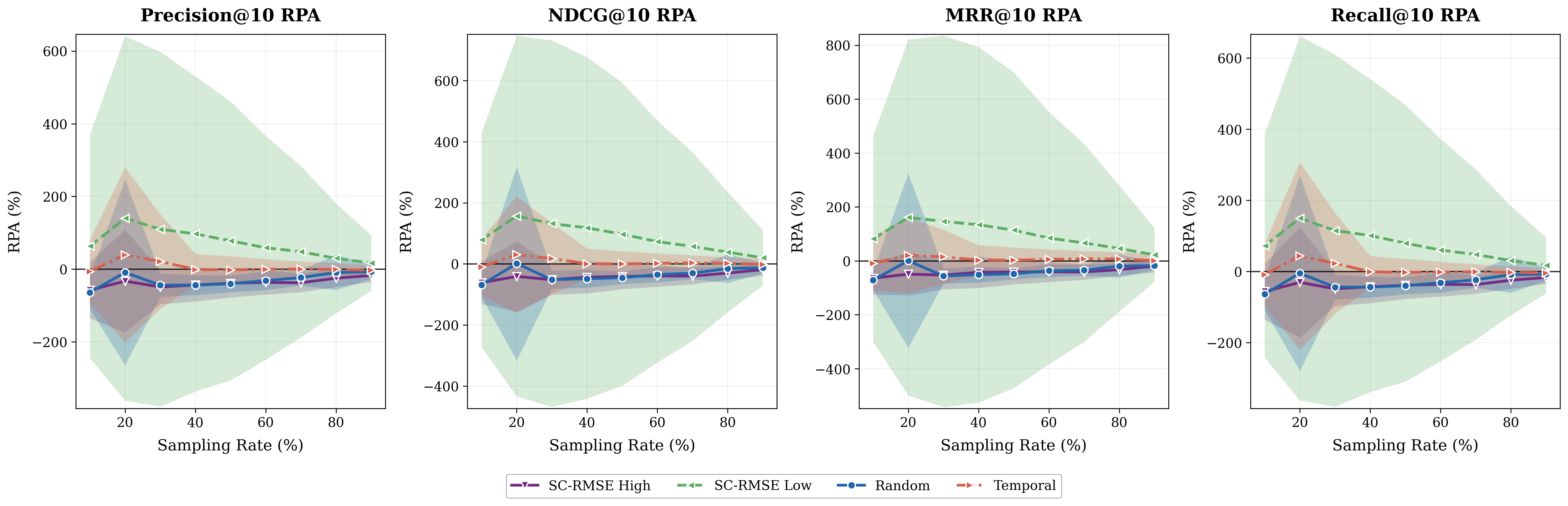}
   \caption{Relative Performance Analysis for EASE algorithm across different sampling strategies. Shaded areas represent the standard deviation of the mean.}
   \label{fig:ease_rpa}
\end{figure*}

\begin{figure*}
   \centering
   \includegraphics[width=0.9\textwidth]{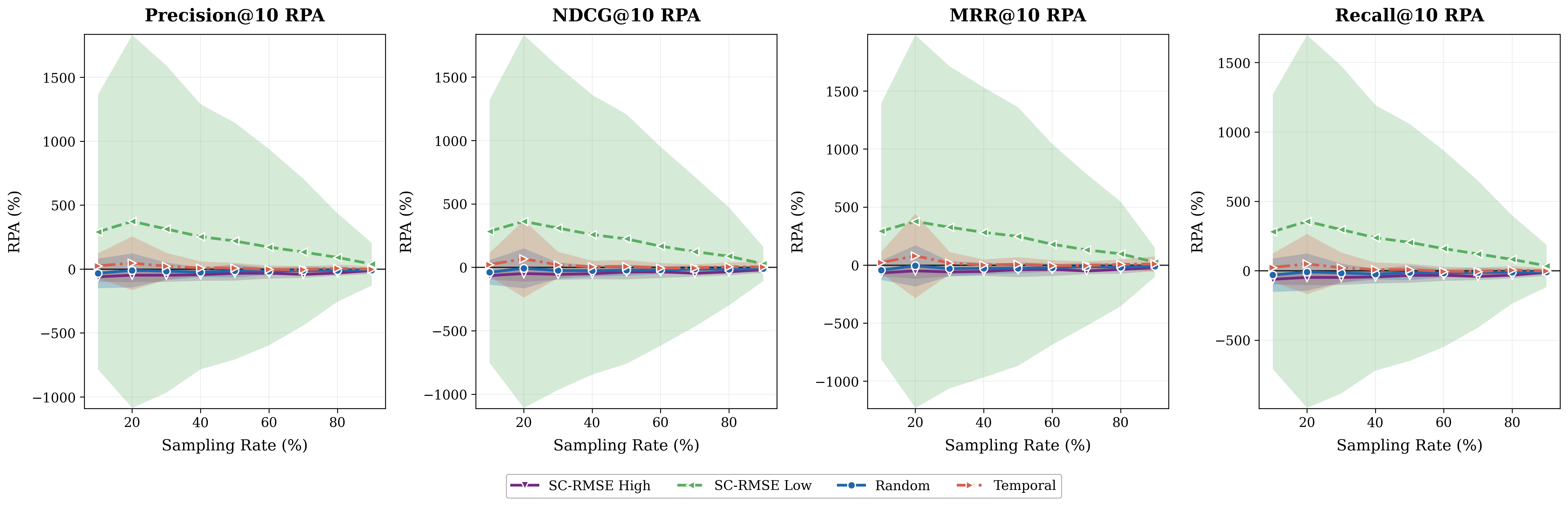}
   \caption{Relative Performance Analysis for MultiDAE algorithm across different sampling strategies. Shaded areas represent the standard deviation of the mean.}
   \label{fig:MultiDAE_rpa}
\end{figure*}

To validate the generalizability of our structural complexity-based sampling approach, we evaluated the same sampling strategies across three additional recommendation algorithms: BPR (Bayesian Personalized Ranking), EASE (Embarrassingly Shallow Autoencoders), and NeuMF (Neural Matrix Factorization).

Across all three algorithms, the relative performance hierarchy remains consistent with our LightGCN findings: SC-Low sampling consistently achieves the highest RPA values, followed by temporal sampling near baseline performance, random sampling with negative performance, and SC RMSE-High sampling with the poorest performance.

While the rankings remain consistent, the magnitude of improvements varies significantly by algorithm architecture. BPR shows moderate improvements with SC-Low achieving approximately 40-50\% RPA at low sampling rates. EASE demonstrates stronger benefits, reaching 60-80\% RPA improvements. NeuMF exhibits the highest gains, with SC-Low sampling achieving 150-200\% RPA for Precision@10 and NDCG@10, and up to 300-400\% RPA for MAP@10 at 10\% sampling rates.

\clearpage
\onecolumn


\begin{table}[ht]
\centering
\caption{Performance Results: Precision@10}
\label{tab:precision@10}
\resizebox{\textwidth}{!}{%
\begin{tabular}{lccccccccccccccccccccccc}
\toprule
\textbf{Model} & \textbf{Amazon} & \textbf{Amazon} & \textbf{Amazon} & \textbf{Anime} & \textbf{Beer} & \textbf{Book} & \textbf{Dian} & \textbf{Doub} & \textbf{Epin} & \textbf{Food} & \textbf{Good} & \textbf{Jest} & \textbf{KDD} & \textbf{Last} & \textbf{ML} & \textbf{Mod} & \textbf{Net} & \textbf{Rate} & \textbf{Rent} & \textbf{Stm} & \textbf{Twitch} & \textbf{Yahoo} & \textbf{Yelp} \\
& \textbf{Books} & \textbf{Music} & \textbf{Movies} & & \textbf{Adv} & \textbf{Cross} & \textbf{ping} & \textbf{an} & \textbf{ions} & & \textbf{reads} & \textbf{er} & \textbf{2010} & \textbf{fm} & \textbf{1M} & \textbf{cloth} & \textbf{flix} & \textbf{beer} & \textbf{Runway} & \textbf{eam} & \textbf{100K} & \textbf{Music} & \\
\midrule
\multicolumn{24}{l}{\textit{Traditional}} \\
Pop & 0.0218 & 0.0398 & 0.0358 & 0.1451 & 0.0475 & 0.0390 & 0.0106 & 0.0500 & 0.0277 & 0.0362 & 0.1513 & 0.2373 & 0.1309 & 0.1955 & 0.1484 & 0.0822 & 0.1301 & 0.0329 & 0.0518 & 0.0375 & 0.0344 & 0.1324 & 0.0192 \\
ItemKNN & 0.0193 & 0.0333 & 0.0306 & 0.2411 & 0.0217 & 0.0213 & 0.0235 & 0.0645 & 0.0175 & 0.0211 & 0.2288 & 0.3497 & 0.4995 & 0.2465 & 0.2542 & 0.0569 & 0.2290 & 0.0218 & 0.0377 & 0.0212 & 0.0604 & 0.1480 & 0.0345 \\
\midrule
\multicolumn{24}{l}{\textit{Matrix Factorization}} \\
BPR & 0.0195 & 0.0368 & 0.0266 & 0.1848 & 0.0474 & 0.0294 & 0.0199 & 0.0538 & 0.0145 & 0.0173 & 0.1896 & 0.3473 & 0.3251 & 0.2487 & 0.2028 & 0.0587 & 0.1819 & 0.0310 & 0.0280 & 0.0121 & 0.0386 & 0.1414 & 0.0389 \\
FISM & 0.0229 & 0.0407 & 0.0374 & 0.1463 & 0.0476 & 0.0394 & 0.0102 & 0.0633 & 0.0302 & 0.0368 & 0.1490 & 0.3500 & 0.1326 & 0.1949 & 0.1503 & 0.0825 & 0.1311 & 0.0330 & \textbf{0.0519} & 0.0402 & 0.0348 & 0.1335 & 0.0195 \\
DMF & 0.0202 & 0.0395 & 0.0364 & 0.1605 & 0.0470 & 0.0396 & 0.0196 & 0.0571 & 0.0285 & 0.0367 & 0.1678 & 0.3458 & 0.2743 & 0.1931 & 0.1716 & 0.0818 & 0.1711 & 0.0322 & 0.0425 & 0.0406 & 0.0372 & 0.1354 & 0.0322 \\
ENMF & 0.0158 & 0.0276 & 0.0220 & 0.0140 & 0.0118 & 0.0143 & 0.0116 & 0.0488 & 0.0138 & 0.0155 & 0.0121 & 0.1745 & 0.0326 & 0.0442 & 0.0153 & 0.0508 & 0.0190 & 0.0119 & 0.0302 & 0.0107 & 0.0098 & 0.0342 & 0.0174 \\
\midrule
\multicolumn{24}{l}{\textit{Neural Networks}} \\
NeuMF & 0.0208 & 0.0371 & 0.0353 & 0.2098 & 0.0482 & 0.0381 & 0.0152 & 0.0636 & 0.0277 & 0.0362 & 0.1960 & 0.3482 & 0.3056 & 0.2781 & 0.2314 & 0.0826 & 0.2088 & 0.0317 & 0.0497 & 0.0397 & 0.0372 & 0.1518 & 0.0431 \\
NNCF & 0.0208 & 0.0379 & 0.0336 & 0.2000 & 0.0453 & 0.0343 & 0.0118 & 0.0573 & 0.0271 & 0.0313 & 0.1649 & 0.3464 & 0.3165 & 0.2490 & 0.2173 & 0.0816 & 0.1976 & 0.0302 & 0.0399 & 0.0322 & 0.0277 & 0.1447 & 0.0431 \\
ConvNCF & 0.0148 & 0.0336 & 0.0302 & 0.1378 & 0.0438 & 0.0332 & 0.0097 & 0.0627 & 0.0161 & 0.0271 & 0.1301 & 0.3348 & 0.1038 & 0.1815 & 0.1410 & 0.0806 & 0.1246 & 0.0264 & 0.0501 & 0.0200 & 0.0140 & 0.1288 & 0.0162 \\
CDAE & 0.0185 & 0.0374 & 0.0327 & 0.1329 & 0.0137 & 0.0303 & 0.0102 & 0.0613 & 0.0135 & 0.0248 & 0.0107 & 0.3353 & 0.0864 & 0.0091 & 0.1406 & 0.0823 & 0.0888 & 0.0116 & \textbf{0.0519} & 0.0102 & 0.0108 & 0.1294 & 0.0176 \\
EASE & 0.0195 & 0.0286 & 0.0247 & \textbf{0.2568} & 0.0415 & 0.0231 & 0.0296 & 0.0657 & 0.0196 & 0.0162 & \textbf{0.2414} & 0.3498 & \textbf{0.5041} & \textbf{0.3061} & \textbf{0.2717} & 0.0692 & \textbf{0.2389} & 0.0319 & 0.0208 & 0.0249 & 0.0753 & \textbf{0.1531} & 0.0372 \\
\midrule
\multicolumn{24}{l}{\textit{Graph Neural Networks}} \\
GCMC & 0.0172 & 0.0405 & 0.0353 & 0.1751 & 0.0501 & 0.0364 & 0.0267 & 0.0628 & 0.0156 & 0.0350 & 0.1702 & 0.3501 & 0.2739 & 0.2361 & 0.1103 & 0.0012 & 0.1519 & 0.0333 & 0.0072 & 0.0373 & 0.0395 & 0.1431 & 0.0464 \\
SpectralCF & 0.0228 & 0.0399 & 0.0364 & 0.1409 & 0.0452 & 0.0386 & 0.0100 & 0.0625 & 0.0289 & 0.0361 & 0.1455 & 0.3472 & 0.1787 & 0.1901 & 0.1436 & 0.0820 & 0.1293 & 0.0302 & 0.0509 & 0.0396 & 0.0342 & 0.1304 & 0.0195 \\
NGCF & 0.0255 & 0.0388 & 0.0314 & 0.1876 & 0.0389 & 0.0287 & 0.0445 & 0.0577 & 0.0194 & 0.0189 & 0.1550 & 0.3460 & 0.1671 & 0.2700 & 0.2009 & 0.0779 & 0.1867 & 0.0248 & 0.0437 & 0.0183 & 0.0650 & 0.1441 & 0.0493 \\
DGCF & 0.0259 & 0.0424 & 0.0350 & 0.1806 & \textbf{0.0526} & 0.0352 & 0.0425 & --- & 0.0187 & 0.0230 & 0.1968 & 0.3500 & 0.3019 & 0.2541 & 0.1952 & 0.0588 & 0.1651 & \textbf{0.0361} & 0.0299 & 0.0173 & 0.0656 & 0.1390 & 0.0503 \\
LightGCN & 0.0276 & 0.0433 & 0.0380 & 0.1654 & 0.0522 & 0.0400 & 0.0487 & 0.0616 & 0.0209 & 0.0298 & 0.1953 & 0.3497 & 0.2707 & 0.2523 & 0.1722 & 0.0655 & 0.1382 & 0.0348 & 0.0326 & 0.0202 & 0.0724 & 0.1375 & 0.0508 \\
SGL & \textbf{0.0284} & \textbf{0.0440} & 0.0367 & 0.1872 & 0.0312 & 0.0328 & \textbf{0.0489} & 0.0564 & 0.0225 & 0.0221 & 0.2078 & 0.3473 & 0.3353 & 0.2785 & 0.2175 & 0.0751 & 0.1936 & 0.0231 & 0.0431 & 0.0139 & \textbf{0.0808} & 0.1429 & \textbf{0.0553} \\
SimpleX & 0.0222 & 0.0353 & 0.0274 & 0.1952 & 0.0303 & 0.0219 & 0.0379 & 0.0609 & 0.0162 & 0.0168 & 0.1772 & 0.3020 & 0.3735 & 0.2645 & 0.1950 & 0.0582 & 0.1849 & 0.0205 & 0.0235 & 0.0146 & 0.0617 & 0.1465 & 0.0425 \\
\midrule
\multicolumn{24}{l}{\textit{Generative / VAE}} \\
MultiDAE & 0.0239 & 0.0425 & \textbf{0.0386} & 0.1468 & 0.0485 & \textbf{0.0402} & 0.0338 & 0.0667 & 0.0302 & \textbf{0.0377} & 0.1256 & \textbf{0.3511} & 0.1245 & 0.2203 & 0.1521 & 0.0804 & 0.1320 & 0.0325 & 0.0453 & \textbf{0.0411} & 0.0399 & 0.1427 & 0.0458 \\
MultiVAE & 0.0238 & 0.0424 & 0.0382 & 0.1493 & 0.0485 & 0.0400 & 0.0334 & 0.0668 & 0.0296 & 0.0375 & 0.1345 & 0.3498 & 0.1618 & 0.2194 & 0.1532 & 0.0809 & 0.1340 & 0.0322 & 0.0467 & 0.0410 & 0.0391 & 0.1434 & 0.0463 \\
MacridVAE & 0.0230 & 0.0380 & 0.0308 & 0.1461 & 0.0366 & 0.0294 & 0.0231 & 0.0662 & 0.0189 & 0.0267 & 0.1310 & 0.3401 & 0.2042 & 0.2730 & 0.1335 & 0.0819 & 0.1219 & 0.0230 & 0.0470 & 0.0358 & 0.0362 & 0.1473 & 0.0458 \\
RecVAE & 0.0227 & 0.0362 & 0.0371 & 0.1440 & 0.0477 & 0.0397 & 0.0220 & \textbf{0.0673} & \textbf{0.0308} & 0.0372 & 0.1114 & 0.3457 & 0.1094 & 0.2154 & 0.1507 & \textbf{0.0828} & 0.1339 & 0.0313 & 0.0474 & 0.0354 & 0.0374 & 0.1482 & 0.0494 \\
\midrule
\multicolumn{24}{l}{\textit{Others}} \\
NCE-PLRec & 0.0191 & 0.0263 & 0.0229 & 0.1446 & 0.0275 & 0.0209 & 0.0256 & 0.0593 & 0.0173 & 0.0170 & 0.2069 & 0.3428 & 0.4813 & 0.2485 & 0.1199 & 0.0300 & 0.1026 & 0.0253 & 0.0190 & 0.0178 & 0.0488 & 0.0875 & 0.0318 \\
\bottomrule
\end{tabular}
}
\end{table}


\begin{table}[ht]
\centering
\caption{Performance Results: Recall@10}
\label{tab:recall@10}
\resizebox{\textwidth}{!}{%
\begin{tabular}{lccccccccccccccccccccccc}
\toprule
\textbf{Model} & \textbf{Amazon} & \textbf{Amazon} & \textbf{Amazon} & \textbf{Anime} & \textbf{Beer} & \textbf{Book} & \textbf{Dian} & \textbf{Doub} & \textbf{Epin} & \textbf{Food} & \textbf{Good} & \textbf{Jest} & \textbf{KDD} & \textbf{Last} & \textbf{ML} & \textbf{Mod} & \textbf{Net} & \textbf{Rate} & \textbf{Rent} & \textbf{Stm} & \textbf{Twitch} & \textbf{Yahoo} & \textbf{Yelp} \\
& \textbf{Books} & \textbf{Music} & \textbf{Movies} & & \textbf{Adv} & \textbf{Cross} & \textbf{ping} & \textbf{an} & \textbf{ions} & & \textbf{reads} & \textbf{er} & \textbf{2010} & \textbf{fm} & \textbf{1M} & \textbf{cloth} & \textbf{flix} & \textbf{beer} & \textbf{Runway} & \textbf{eam} & \textbf{100K} & \textbf{Music} & \\
\midrule
\multicolumn{24}{l}{\textit{Traditional}} \\
Pop & 0.2125 & 0.3884 & 0.3538 & 0.3628 & 0.1715 & 0.3222 & 0.0975 & 0.4804 & 0.2745 & 0.3446 & 0.2512 & 0.5022 & 0.0869 & 0.4001 & 0.2132 & 0.8215 & 0.2221 & 0.0440 & 0.5173 & 0.3567 & 0.1201 & 0.7035 & 0.1877 \\
ItemKNN & 0.1781 & 0.3214 & 0.2984 & 0.5066 & 0.0771 & 0.1994 & 0.1829 & 0.6251 & 0.1696 & 0.2056 & 0.3298 & 0.7858 & 0.2072 & 0.5031 & 0.3568 & 0.5685 & 0.3613 & 0.0297 & 0.3756 & 0.2044 & 0.2298 & 0.7173 & 0.3371 \\
\midrule
\multicolumn{24}{l}{\textit{Matrix Factorization}} \\
BPR & 0.1853 & 0.3544 & 0.2610 & 0.4194 & 0.1702 & 0.2355 & 0.1708 & 0.5178 & 0.1433 & 0.1640 & 0.2933 & 0.7807 & 0.1317 & 0.5078 & 0.2765 & 0.5861 & 0.2791 & 0.0438 & 0.2793 & 0.1148 & 0.1355 & 0.7253 & 0.3814 \\
FISM & 0.2234 & 0.3975 & 0.3689 & 0.3699 & 0.1706 & 0.3264 & 0.0932 & 0.6124 & 0.2996 & 0.3509 & 0.2538 & 0.7863 & 0.0878 & 0.3992 & 0.2115 & 0.8238 & 0.2194 & 0.0451 & \textbf{0.5188} & 0.3819 & 0.1229 & 0.7043 & 0.1907 \\
DMF & 0.1955 & 0.3854 & 0.3595 & 0.3916 & 0.1698 & 0.3287 & 0.1761 & 0.5515 & 0.2819 & 0.3505 & 0.2695 & 0.7787 & 0.1268 & 0.3957 & 0.2358 & 0.8171 & 0.2792 & 0.0450 & 0.4240 & 0.3845 & 0.1315 & 0.7139 & 0.3152 \\
ENMF & 0.1525 & 0.2647 & 0.2161 & 0.0446 & 0.0439 & 0.1243 & 0.1053 & 0.4694 & 0.1363 & 0.1461 & 0.0302 & 0.2864 & 0.0309 & 0.0913 & 0.0261 & 0.5076 & 0.0424 & 0.0154 & 0.3007 & 0.1048 & 0.0406 & 0.1959 & 0.1696 \\
\midrule
\multicolumn{24}{l}{\textit{Neural Networks}} \\
NeuMF & 0.1996 & 0.3604 & 0.3482 & 0.4697 & 0.1772 & 0.3171 & 0.1273 & 0.6154 & 0.2740 & 0.3443 & 0.3105 & 0.7827 & 0.1510 & 0.5668 & 0.3391 & 0.8255 & 0.3366 & 0.0454 & 0.4953 & 0.3764 & 0.1298 & \textbf{0.7659} & 0.4227 \\
NNCF & 0.2011 & 0.3689 & 0.3317 & 0.4566 & 0.1631 & 0.2868 & 0.1069 & 0.5527 & 0.2682 & 0.2967 & 0.2795 & 0.7782 & 0.1362 & 0.5082 & 0.3012 & 0.8155 & 0.3126 & 0.0418 & 0.3978 & 0.3083 & 0.0990 & 0.7418 & 0.4245 \\
ConvNCF & 0.1451 & 0.3285 & 0.2986 & 0.3471 & 0.1572 & 0.2756 & 0.0904 & 0.6069 & 0.1588 & 0.2581 & 0.2206 & 0.7394 & 0.0777 & 0.3719 & 0.1953 & 0.8052 & 0.2061 & 0.0350 & 0.5001 & 0.1926 & 0.0544 & 0.6876 & 0.1583 \\
CDAE & 0.1800 & 0.3642 & 0.3226 & 0.2991 & 0.0505 & 0.2487 & 0.0945 & 0.5927 & 0.1346 & 0.2357 & 0.0288 & 0.7449 & 0.0423 & 0.0191 & 0.1911 & 0.8226 & 0.1339 & 0.0147 & 0.5181 & 0.1000 & 0.0462 & 0.6754 & 0.1718 \\
EASE & 0.1794 & 0.2714 & 0.2380 & \textbf{0.5115} & 0.1439 & 0.1747 & 0.2387 & 0.6364 & 0.1909 & 0.1497 & \textbf{0.3512} & 0.7857 & \textbf{0.2233} & \textbf{0.6219} & \textbf{0.3838} & 0.6910 & \textbf{0.3669} & 0.0433 & 0.2056 & 0.2280 & 0.2754 & 0.7295 & 0.3627 \\
\midrule
\multicolumn{24}{l}{\textit{Graph Neural Networks}} \\
GCMC & 0.1682 & 0.3932 & 0.3490 & 0.4177 & 0.1827 & 0.3011 & 0.2406 & 0.6071 & 0.1536 & 0.3328 & 0.2807 & \textbf{0.7867} & 0.1276 & 0.4823 & 0.1575 & 0.0116 & 0.2572 & 0.0471 & 0.0720 & 0.3528 & 0.1395 & 0.7367 & 0.4566 \\
SpectralCF & 0.2214 & 0.3898 & 0.3597 & 0.3548 & 0.1627 & 0.3207 & 0.0919 & 0.6044 & 0.2869 & 0.3438 & 0.2497 & 0.7795 & 0.1082 & 0.3893 & 0.2042 & 0.8191 & 0.2134 & 0.0418 & 0.5084 & 0.3763 & 0.1199 & 0.6971 & 0.1906 \\
NGCF & 0.2430 & 0.3749 & 0.3078 & 0.4345 & 0.1435 & 0.2392 & 0.3861 & 0.5564 & 0.1908 & 0.1798 & 0.2770 & 0.7778 & 0.1221 & 0.5507 & 0.2906 & 0.7782 & 0.3003 & 0.0367 & 0.4357 & 0.1732 & 0.2428 & 0.7394 & 0.4852 \\
DGCF & 0.2451 & 0.4096 & 0.3416 & 0.4181 & \textbf{0.1891} & 0.2836 & 0.3637 & --- & 0.1828 & 0.2161 & 0.3109 & 0.7865 & 0.1483 & 0.5191 & 0.2704 & 0.5872 & 0.2620 & \textbf{0.0515} & 0.2980 & 0.1596 & 0.2370 & 0.7215 & 0.4953 \\
LightGCN & 0.2620 & 0.4192 & 0.3712 & 0.3956 & 0.1883 & 0.3262 & \textbf{0.4219} & 0.5954 & 0.2043 & 0.2815 & 0.3101 & 0.7863 & 0.1287 & 0.5153 & 0.2392 & 0.6542 & 0.2293 & 0.0492 & 0.3238 & 0.1863 & 0.2664 & 0.7176 & 0.5007 \\
SGL & \textbf{0.2695} & \textbf{0.4242} & 0.3580 & 0.3427 & 0.1075 & 0.2579 & 0.4217 & 0.5431 & 0.2201 & 0.2065 & 0.2933 & 0.7824 & 0.1213 & 0.5677 & 0.2693 & 0.7505 & 0.2817 & 0.0324 & 0.4292 & 0.1273 & \textbf{0.2977} & 0.7312 & \textbf{0.5442} \\
SimpleX & 0.2099 & 0.3390 & 0.2686 & 0.4402 & 0.1062 & 0.1853 & 0.3233 & 0.5878 & 0.1590 & 0.1604 & 0.2893 & 0.6888 & 0.1753 & 0.5394 & 0.2773 & 0.5816 & 0.2961 & 0.0264 & 0.2334 & 0.1384 & 0.2245 & 0.7387 & 0.4182 \\
\midrule
\multicolumn{24}{l}{\textit{Generative / VAE}} \\
MultiDAE & 0.2302 & 0.4113 & \textbf{0.3803} & 0.3676 & 0.1752 & \textbf{0.3326} & 0.2902 & 0.6457 & 0.2991 & \textbf{0.3596} & 0.2436 & \textbf{0.7867} & 0.0790 & 0.4503 & 0.2140 & 0.8036 & 0.2181 & 0.0436 & 0.4507 & \textbf{0.3900} & 0.1439 & 0.7329 & 0.4504 \\
MultiVAE & 0.2288 & 0.4101 & 0.3768 & 0.3665 & 0.1729 & 0.3320 & 0.2874 & 0.6472 & 0.2934 & 0.3585 & 0.2473 & 0.7864 & 0.0960 & 0.4498 & 0.2182 & 0.8081 & 0.2237 & 0.0447 & 0.4652 & 0.3880 & 0.1421 & 0.7349 & 0.4540 \\
MacridVAE & 0.2172 & 0.3653 & 0.3028 & 0.3738 & 0.1322 & 0.2530 & 0.1914 & 0.6413 & 0.1876 & 0.2579 & 0.2472 & 0.7594 & 0.1237 & 0.5565 & 0.1966 & 0.8184 & 0.2098 & 0.0309 & 0.4684 & 0.3394 & 0.1295 & 0.7497 & 0.4497 \\
RecVAE & 0.2211 & 0.3485 & 0.3664 & 0.3620 & 0.1717 & 0.3292 & 0.1884 & \textbf{0.6521} & \textbf{0.3054} & 0.3548 & 0.2253 & 0.7787 & 0.0971 & 0.4398 & 0.2138 & \textbf{0.8277} & 0.2296 & 0.0427 & 0.4721 & 0.3346 & 0.1323 & 0.7494 & 0.4851 \\
\midrule
\multicolumn{24}{l}{\textit{Others}} \\
NCE-PLRec & 0.1785 & 0.2541 & 0.2227 & 0.3788 & 0.0977 & 0.1768 & 0.2126 & 0.5716 & 0.1694 & 0.1624 & 0.3167 & 0.7689 & 0.2197 & 0.5064 & 0.2344 & 0.3003 & 0.2255 & 0.0333 & 0.1881 & 0.1668 & 0.1844 & 0.4863 & 0.3120 \\
\bottomrule
\end{tabular}
}
\end{table}


\begin{table}[ht]
\centering
\caption{Performance Results: NDCG@10}
\label{tab:ndcg@10}
\resizebox{\textwidth}{!}{%
\begin{tabular}{lccccccccccccccccccccccc}
\toprule
\textbf{Model} & \textbf{Amazon} & \textbf{Amazon} & \textbf{Amazon} & \textbf{Anime} & \textbf{Beer} & \textbf{Book} & \textbf{Dian} & \textbf{Doub} & \textbf{Epin} & \textbf{Food} & \textbf{Good} & \textbf{Jest} & \textbf{KDD} & \textbf{Last} & \textbf{ML} & \textbf{Mod} & \textbf{Net} & \textbf{Rate} & \textbf{Rent} & \textbf{Stm} & \textbf{Twitch} & \textbf{Yahoo} & \textbf{Yelp} \\
& \textbf{Books} & \textbf{Music} & \textbf{Movies} & & \textbf{Adv} & \textbf{Cross} & \textbf{ping} & \textbf{an} & \textbf{ions} & & \textbf{reads} & \textbf{er} & \textbf{2010} & \textbf{fm} & \textbf{1M} & \textbf{cloth} & \textbf{flix} & \textbf{beer} & \textbf{Runway} & \textbf{eam} & \textbf{100K} & \textbf{Music} & \\
\midrule
\multicolumn{24}{l}{\textit{Traditional}} \\
Pop & 0.1333 & 0.2370 & 0.2369 & 0.3205 & 0.1143 & 0.1992 & 0.0452 & 0.2248 & 0.1607 & 0.2078 & 0.2755 & 0.3646 & 0.1750 & 0.3668 & 0.2313 & 0.4704 & 0.2160 & 0.0414 & 0.3159 & 0.2262 & 0.0895 & 0.5034 & 0.0940 \\
ItemKNN & 0.1120 & 0.1762 & 0.1553 & 0.5044 & 0.0500 & 0.0889 & 0.1178 & 0.3190 & 0.0891 & 0.0882 & 0.3945 & 0.6622 & 0.5973 & 0.5106 & 0.3975 & 0.3245 & 0.3791 & 0.0321 & 0.1657 & 0.0924 & 0.1797 & 0.5854 & 0.1899 \\
\midrule
\multicolumn{24}{l}{\textit{Matrix Factorization}} \\
BPR & 0.1205 & 0.2382 & 0.1773 & 0.3873 & 0.1141 & 0.1421 & 0.1006 & 0.2671 & 0.0841 & 0.0907 & 0.3292 & 0.6545 & 0.3986 & 0.4891 & 0.3142 & 0.3336 & 0.2876 & 0.0413 & 0.1653 & 0.0569 & 0.1048 & 0.5428 & 0.2035 \\
FISM & 0.1384 & 0.2410 & 0.2442 & 0.3238 & 0.1128 & 0.2000 & 0.0434 & 0.3103 & 0.1736 & 0.2109 & 0.2727 & 0.6626 & 0.1729 & 0.3662 & 0.2328 & 0.4690 & 0.2132 & 0.0418 & 0.3161 & 0.2435 & 0.0910 & 0.5020 & 0.0957 \\
DMF & 0.1169 & 0.2288 & 0.2324 & 0.3484 & 0.1111 & 0.1987 & 0.0865 & 0.2777 & 0.1645 & 0.2082 & 0.2941 & 0.6627 & 0.3345 & 0.3622 & 0.2659 & 0.4582 & 0.2780 & 0.0418 & 0.2577 & \textbf{0.2452} & 0.0961 & 0.5177 & 0.1587 \\
ENMF & 0.0894 & 0.1698 & 0.1403 & 0.0290 & 0.0256 & 0.0641 & 0.0511 & 0.2212 & 0.0749 & 0.0751 & 0.0201 & 0.2306 & 0.0458 & 0.0813 & 0.0230 & 0.3014 & 0.0323 & 0.0146 & 0.1886 & 0.0493 & 0.0235 & 0.1174 & 0.0895 \\
\midrule
\multicolumn{24}{l}{\textit{Neural Networks}} \\
NeuMF & 0.1240 & 0.2252 & 0.2306 & 0.4329 & 0.1153 & 0.1924 & 0.0670 & 0.3141 & 0.1604 & 0.2045 & 0.3379 & 0.6572 & 0.3807 & 0.5454 & 0.3632 & 0.4706 & 0.3390 & 0.0407 & 0.3020 & 0.2375 & 0.0982 & 0.5875 & 0.2277 \\
NNCF & 0.1183 & 0.2237 & 0.2149 & 0.4140 & 0.1058 & 0.1714 & 0.0521 & 0.2797 & 0.1485 & 0.1818 & 0.3006 & 0.6544 & 0.3909 & 0.4831 & 0.3371 & 0.4596 & 0.3175 & 0.0395 & 0.2214 & 0.1794 & 0.0654 & 0.5583 & 0.2155 \\
ConvNCF & 0.0822 & 0.1946 & 0.1951 & 0.3001 & 0.1034 & 0.1658 & 0.0411 & 0.3057 & 0.0809 & 0.1502 & 0.2351 & 0.6007 & 0.1438 & 0.3423 & 0.2142 & 0.4521 & 0.1977 & 0.0327 & 0.3028 & 0.1089 & 0.0349 & 0.4827 & 0.0781 \\
CDAE & 0.1114 & 0.2183 & 0.2219 & 0.2863 & 0.0307 & 0.1556 & 0.0440 & 0.3064 & 0.0688 & 0.1417 & 0.0201 & 0.6045 & 0.1061 & 0.0140 & 0.2187 & 0.4718 & 0.1371 & 0.0142 & \textbf{0.3168} & 0.0443 & 0.0267 & 0.4812 & 0.0868 \\
EASE & 0.1270 & 0.1909 & 0.1635 & \textbf{0.5271} & 0.0981 & 0.1081 & 0.1592 & 0.3259 & 0.1106 & 0.0826 & \textbf{0.4194} & 0.6662 & \textbf{0.6213} & \textbf{0.6114} & \textbf{0.4353} & 0.4278 & \textbf{0.3986} & 0.0471 & 0.1266 & 0.1378 & 0.2263 & \textbf{0.5998} & 0.2066 \\
\midrule
\multicolumn{24}{l}{\textit{Graph Neural Networks}} \\
GCMC & 0.0984 & 0.2501 & 0.2346 & 0.3772 & 0.1201 & 0.1830 & 0.1186 & 0.3064 & 0.0681 & 0.1983 & 0.3044 & 0.6667 & 0.3350 & 0.4552 & 0.1731 & 0.0061 & 0.2530 & 0.0424 & 0.0386 & 0.2319 & 0.1012 & 0.5639 & 0.2333 \\
SpectralCF & 0.1367 & 0.2343 & 0.2385 & 0.3108 & 0.1063 & 0.1948 & 0.0434 & 0.3124 & 0.1662 & 0.2055 & 0.2647 & 0.6527 & 0.2334 & 0.3601 & 0.2204 & 0.4649 & 0.2070 & 0.0394 & 0.3095 & 0.2332 & 0.0895 & 0.4943 & 0.0954 \\
NGCF & 0.1390 & 0.2258 & 0.1809 & 0.3804 & 0.0893 & 0.1259 & 0.2257 & 0.2861 & 0.1005 & 0.0896 & 0.2723 & 0.6502 & 0.2134 & 0.5032 & 0.2982 & 0.4357 & 0.2928 & 0.0324 & 0.2534 & 0.0880 & 0.1753 & 0.5517 & 0.2546 \\
DGCF & 0.1631 & 0.2743 & 0.2319 & 0.3826 & 0.1250 & 0.1725 & 0.2267 & --- & 0.1106 & 0.1251 & 0.3491 & 0.6663 & 0.3747 & 0.4973 & 0.3047 & 0.3393 & 0.2662 & \textbf{0.0473} & 0.1781 & 0.0978 & 0.1902 & 0.5342 & 0.2703 \\
LightGCN & 0.1731 & 0.2779 & \textbf{0.2496} & 0.3566 & \textbf{0.1251} & 0.1989 & 0.2596 & 0.3095 & 0.1240 & 0.1673 & 0.3484 & \textbf{0.6676} & 0.3326 & 0.4930 & 0.2678 & 0.3788 & 0.2250 & 0.0457 & 0.1857 & 0.1163 & 0.2109 & 0.5303 & 0.2759 \\
SGL & \textbf{0.1782} & \textbf{0.2923} & 0.2440 & 0.3498 & 0.0722 & 0.1573 & \textbf{0.2652} & 0.2874 & 0.1343 & 0.1173 & 0.3403 & 0.6655 & 0.3919 & 0.5523 & 0.3251 & 0.4333 & 0.3124 & 0.0299 & 0.2547 & 0.0706 & \textbf{0.2386} & 0.5655 & \textbf{0.3048} \\
SimpleX & 0.1276 & 0.2195 & 0.1623 & 0.4134 & 0.0713 & 0.0981 & 0.1988 & 0.3053 & 0.0826 & 0.0834 & 0.3166 & 0.5984 & 0.4717 & 0.5169 & 0.3025 & 0.3324 & 0.3071 & 0.0260 & 0.1222 & 0.0739 & 0.1732 & 0.5682 & 0.2298 \\
\midrule
\multicolumn{24}{l}{\textit{Generative / VAE}} \\
MultiDAE & 0.1421 & 0.2665 & 0.2493 & 0.3256 & 0.1144 & \textbf{0.2020} & 0.1697 & 0.3397 & 0.1723 & \textbf{0.2136} & 0.2230 & 0.6671 & 0.1570 & 0.4282 & 0.2363 & 0.4807 & 0.2112 & 0.0419 & 0.2631 & 0.2428 & 0.1022 & 0.5519 & 0.2471 \\
MultiVAE & 0.1407 & 0.2650 & 0.2481 & 0.3268 & 0.1146 & 0.2020 & 0.1691 & 0.3403 & 0.1712 & 0.2118 & 0.2369 & 0.6659 & 0.2004 & 0.4266 & 0.2389 & 0.4828 & 0.2165 & 0.0422 & 0.2759 & 0.2429 & 0.1012 & 0.5557 & 0.2497 \\
MacridVAE & 0.1353 & 0.2374 & 0.1946 & 0.3195 & 0.0866 & 0.1434 & 0.1151 & 0.3384 & 0.0954 & 0.1404 & 0.2435 & 0.6316 & 0.2713 & 0.5441 & 0.2066 & 0.4813 & 0.2016 & 0.0292 & 0.2806 & 0.2104 & 0.0928 & 0.5818 & 0.2478 \\
RecVAE & 0.1371 & 0.2276 & 0.2420 & 0.3153 & 0.1129 & 0.2011 & 0.1100 & \textbf{0.3439} & \textbf{0.1745} & 0.2118 & 0.2006 & 0.6607 & 0.1538 & 0.4219 & 0.2314 & \textbf{0.5017} & 0.2203 & 0.0388 & 0.2845 & 0.2139 & 0.0961 & 0.5856 & 0.2697 \\
\midrule
\multicolumn{24}{l}{\textit{Others}} \\
NCE-PLRec & 0.1225 & 0.1759 & 0.1481 & 0.3639 & 0.0678 & 0.1009 & 0.1413 & 0.2894 & 0.0995 & 0.0841 & 0.3694 & 0.6553 & 0.6008 & 0.5137 & 0.2270 & 0.1382 & 0.2077 & 0.0388 & 0.0962 & 0.0891 & 0.1572 & 0.3677 & 0.1858 \\
\bottomrule
\end{tabular}
}
\end{table}


\begin{table}[ht]
\centering
\caption{Performance Results: MRR@10}
\label{tab:mrr@10}
\resizebox{\textwidth}{!}{%
\begin{tabular}{lccccccccccccccccccccccc}
\toprule
\textbf{Model} & \textbf{Amazon} & \textbf{Amazon} & \textbf{Amazon} & \textbf{Anime} & \textbf{Beer} & \textbf{Book} & \textbf{Dian} & \textbf{Doub} & \textbf{Epin} & \textbf{Food} & \textbf{Good} & \textbf{Jest} & \textbf{KDD} & \textbf{Last} & \textbf{ML} & \textbf{Mod} & \textbf{Net} & \textbf{Rate} & \textbf{Rent} & \textbf{Stm} & \textbf{Twitch} & \textbf{Yahoo} & \textbf{Yelp} \\
& \textbf{Books} & \textbf{Music} & \textbf{Movies} & & \textbf{Adv} & \textbf{Cross} & \textbf{ping} & \textbf{an} & \textbf{ions} & & \textbf{reads} & \textbf{er} & \textbf{2010} & \textbf{fm} & \textbf{1M} & \textbf{cloth} & \textbf{flix} & \textbf{beer} & \textbf{Runway} & \textbf{eam} & \textbf{100K} & \textbf{Music} & \\
\midrule
\multicolumn{24}{l}{\textit{Traditional}} \\
Pop & 0.1103 & 0.1916 & 0.2016 & 0.4568 & 0.1494 & 0.1751 & 0.0312 & 0.1492 & 0.1267 & 0.1694 & 0.4428 & 0.4290 & 0.2689 & 0.5233 & 0.3911 & 0.3624 & 0.3443 & 0.0951 & 0.2543 & 0.1906 & 0.1195 & 0.5041 & 0.0667 \\
ItemKNN & 0.0947 & 0.1339 & 0.1142 & 0.6783 & 0.0695 & 0.0589 & 0.1083 & 0.2272 & 0.0661 & 0.0545 & 0.6105 & 0.7017 & 0.7324 & 0.7391 & 0.5881 & 0.2502 & 0.5320 & 0.0855 & 0.1034 & 0.0625 & 0.2334 & 0.6169 & 0.1467 \\
\midrule
\multicolumn{24}{l}{\textit{Matrix Factorization}} \\
BPR & 0.1026 & 0.2037 & 0.1527 & 0.5399 & 0.1500 & 0.1252 & 0.0851 & 0.1937 & 0.0667 & 0.0708 & 0.5072 & 0.6953 & 0.5770 & 0.6858 & 0.5071 & 0.2564 & 0.4302 & 0.0960 & 0.1305 & 0.0406 & 0.1414 & 0.5542 & 0.1512 \\
FISM & 0.1135 & 0.1938 & 0.2066 & 0.4612 & 0.1463 & 0.1746 & 0.0302 & 0.2202 & \textbf{0.1359} & 0.1714 & 0.4371 & 0.7013 & 0.2529 & 0.5263 & 0.3943 & 0.3598 & 0.3358 & 0.0947 & 0.2540 & 0.2063 & 0.1205 & 0.5029 & 0.0681 \\
DMF & 0.0941 & 0.1818 & 0.1939 & 0.4854 & 0.1440 & 0.1716 & 0.0631 & 0.1965 & 0.1291 & 0.1678 & 0.4612 & 0.7033 & 0.4807 & 0.5253 & 0.4376 & 0.3482 & 0.4112 & 0.0978 & 0.2066 & \textbf{0.2077} & 0.1257 & 0.5190 & 0.1128 \\
ENMF & 0.0716 & 0.1421 & 0.1179 & 0.0410 & 0.0325 & 0.0501 & 0.0368 & 0.1481 & 0.0569 & 0.0557 & 0.0350 & 0.2990 & 0.1069 & 0.1532 & 0.0434 & 0.2373 & 0.0572 & 0.0354 & 0.1541 & 0.0333 & 0.0271 & 0.1211 & 0.0664 \\
\midrule
\multicolumn{24}{l}{\textit{Neural Networks}} \\
NeuMF & 0.1025 & 0.1849 & 0.1952 & 0.5736 & 0.1459 & 0.1666 & 0.0545 & 0.2245 & 0.1262 & 0.1651 & 0.5205 & 0.6960 & 0.5572 & 0.7330 & 0.5486 & 0.3615 & 0.4791 & 0.0903 & 0.2429 & 0.2002 & 0.1318 & 0.5968 & 0.1702 \\
NNCF & 0.0945 & 0.1802 & 0.1798 & 0.5549 & 0.1367 & 0.1476 & 0.0376 & 0.1993 & 0.1129 & 0.1499 & 0.4717 & 0.6984 & 0.5782 & 0.6713 & 0.5270 & 0.3505 & 0.4575 & 0.0897 & 0.1682 & 0.1440 & 0.0804 & 0.5658 & 0.1540 \\
ConvNCF & 0.0641 & 0.1548 & 0.1640 & 0.4226 & 0.1342 & 0.1441 & 0.0277 & 0.2159 & 0.0578 & 0.1203 & 0.3921 & 0.6484 & 0.2439 & 0.5050 & 0.3606 & 0.3440 & 0.3148 & 0.0766 & 0.2423 & 0.0861 & 0.0455 & 0.4818 & 0.0548 \\
CDAE & 0.0916 & 0.1746 & 0.1916 & 0.4375 & 0.0401 & 0.1387 & 0.0303 & 0.2219 & 0.0493 & 0.1161 & 0.0319 & 0.6502 & 0.1640 & 0.0252 & 0.3833 & 0.3638 & 0.2279 & 0.0342 & \textbf{0.2551} & 0.0284 & 0.0309 & 0.4836 & 0.0621 \\
EASE & 0.1140 & 0.1680 & 0.1426 & \textbf{0.7044} & 0.1365 & 0.1005 & 0.1451 & 0.2328 & 0.0874 & 0.0651 & \textbf{0.6448} & 0.7062 & \textbf{0.7585} & \textbf{0.8077} & \textbf{0.6438} & 0.3465 & \textbf{0.5651} & \textbf{0.1244} & 0.1027 & 0.1158 & 0.2987 & \textbf{0.6278} & 0.1611 \\
\midrule
\multicolumn{24}{l}{\textit{Graph Neural Networks}} \\
GCMC & 0.0779 & 0.2072 & 0.2000 & 0.5207 & 0.1552 & 0.1591 & 0.0862 & 0.2168 & 0.0429 & 0.1604 & 0.4817 & 0.7028 & 0.4692 & 0.6325 & 0.2878 & 0.0044 & 0.3865 & 0.0930 & 0.0286 & 0.2000 & 0.1265 & 0.5749 & 0.1673 \\
SpectralCF & 0.1118 & 0.1876 & 0.2020 & 0.4418 & 0.1372 & 0.1691 & 0.0305 & 0.2261 & 0.1302 & 0.1666 & 0.4262 & 0.6945 & 0.3785 & 0.5259 & 0.3695 & 0.3562 & 0.3278 & 0.0920 & 0.2485 & 0.1939 & 0.1208 & 0.4944 & 0.0678 \\
NGCF & 0.1095 & 0.1813 & 0.1433 & 0.4983 & 0.1121 & 0.0988 & 0.1835 & 0.2063 & 0.0741 & 0.0641 & 0.4082 & 0.6897 & 0.3360 & 0.6555 & 0.4457 & 0.3308 & 0.4085 & 0.0724 & 0.1979 & 0.0640 & 0.2185 & 0.5489 & 0.1859 \\
DGCF & 0.1404 & 0.2339 & 0.1997 & 0.5293 & \textbf{0.1603} & 0.1515 & 0.1925 & --- & 0.0894 & 0.1002 & 0.5338 & 0.7059 & 0.5361 & 0.6888 & 0.4948 & 0.2631 & 0.4044 & 0.1071 & 0.1413 & 0.0837 & 0.2507 & 0.5418 & 0.2032 \\
LightGCN & 0.1485 & 0.2356 & \textbf{0.2137} & 0.4979 & 0.1598 & 0.1737 & 0.2174 & 0.2249 & 0.1007 & 0.1357 & 0.5341 & 0.7071 & 0.4768 & 0.6837 & 0.4466 & 0.2949 & 0.3521 & 0.1034 & 0.1441 & 0.0999 & 0.2726 & 0.5372 & 0.2087 \\
SGL & \textbf{0.1528} & \textbf{0.2529} & 0.2107 & 0.5235 & 0.0980 & 0.1398 & \textbf{0.2252} & 0.2120 & 0.1093 & 0.0933 & 0.5282 & \textbf{0.7084} & 0.5592 & 0.7528 & 0.5270 & 0.3363 & 0.4745 & 0.0676 & 0.2017 & 0.0574 & \textbf{0.3051} & 0.5818 & \textbf{0.2333} \\
SimpleX & 0.1051 & 0.1843 & 0.1312 & 0.5673 & 0.0995 & 0.0780 & 0.1687 & 0.2220 & 0.0604 & 0.0617 & 0.4967 & 0.6961 & 0.6745 & 0.7037 & 0.4837 & 0.2567 & 0.4624 & 0.0628 & 0.0890 & 0.0567 & 0.2250 & 0.5796 & 0.1740 \\
\midrule
\multicolumn{24}{l}{\textit{Generative / VAE}} \\
MultiDAE & 0.1168 & 0.2233 & 0.2099 & 0.4676 & 0.1472 & 0.1750 & 0.1404 & 0.2489 & 0.1346 & \textbf{0.1724} & 0.3380 & 0.7070 & 0.2797 & 0.6076 & 0.3987 & 0.3816 & 0.3314 & 0.0986 & 0.2063 & 0.2036 & 0.1286 & 0.5618 & 0.1873 \\
MultiVAE & 0.1154 & 0.2216 & 0.2093 & 0.4678 & 0.1486 & \textbf{0.1754} & 0.1405 & 0.2492 & 0.1346 & 0.1703 & 0.3689 & 0.7055 & 0.3322 & 0.6042 & 0.4031 & 0.3828 & 0.3443 & 0.0992 & 0.2184 & 0.2044 & 0.1284 & 0.5666 & 0.1895 \\
MacridVAE & 0.1129 & 0.2000 & 0.1625 & 0.4495 & 0.1145 & 0.1180 & 0.0997 & 0.2484 & 0.0680 & 0.1066 & 0.3915 & 0.6789 & 0.4466 & 0.7350 & 0.3564 & 0.3776 & 0.3192 & 0.0680 & 0.2235 & 0.1755 & 0.1216 & 0.5948 & 0.1884 \\
RecVAE & 0.1126 & 0.1920 & 0.2042 & 0.4480 & 0.1456 & 0.1747 & 0.0924 & \textbf{0.2523} & 0.1353 & 0.1714 & 0.3026 & 0.7064 & 0.2808 & 0.6038 & 0.3847 & \textbf{0.4012} & 0.3414 & 0.0855 & 0.2271 & 0.1828 & 0.1242 & 0.6019 & 0.2061 \\
\midrule
\multicolumn{24}{l}{\textit{Others}} \\
NCE-PLRec & 0.1079 & 0.1535 & 0.1267 & 0.5253 & 0.1000 & 0.0852 & 0.1290 & 0.2064 & 0.0790 & 0.0619 & 0.5927 & 0.7045 & 0.7550 & 0.7371 & 0.3885 & 0.0903 & 0.3320 & 0.1082 & 0.0689 & 0.0696 & 0.2238 & 0.3950 & 0.1490 \\
\bottomrule
\end{tabular}
}
\end{table}

\end{document}